\documentclass[prb,twocolumn]{revtex4-2}

\usepackage{graphicx}
\usepackage{amsmath, amsfonts,amssymb}
\usepackage{color}
\usepackage[dvipsnames]{xcolor}

\newcommand{\be}{\begin{equation}}
\newcommand{\ee}{\end{equation}}

\newcommand{\bea}{\begin{eqnarray}}
\newcommand{\eea}{\end{eqnarray}}

\newcommand{\Tr}{{\rm Tr}}

\renewcommand{\vec}[1]{{\bf #1}}
\renewcommand{\epsilon}{\varepsilon}

\renewcommand{\mod}{{\rm mod}\ }
\newcommand{\norm}[1]{\lVert#1 \rVert}
\usepackage[dvipsnames]{xcolor}
\definecolor{MyColor}{RGB}{0,0,240}

\newcommand{\llangle}{\langle \! \langle }
\newcommand{\rrangle}{\rangle\!\rangle}
\newcommand{\nhat}{\hat n}

\newcommand{\tU}{\tilde U}
\newcommand{\tH}{\tilde H}
\newcommand{\tPsi}{\tilde{\Psi}}
\renewcommand{\a}{{\alpha _1 \ldots \alpha _k}}
\newcommand{\al}{{\alpha_1 \ldots \alpha_\ell}}
\newcommand{\n}{{\hat n_{\alpha _1} \ldots  \hat n_{\alpha _k}}}

\newcommand{\Psia}{\Psi_\al}

\begin{document}

\title{
Hierarchy of many-body invariants and quantized magnetization in anomalous Floquet insulators}
\author{Frederik Nathan$^1$, Dmitry A. Abanin$^2$, Netanel H. Lindner$^3$, Erez Berg$^{4,5}$, and Mark S. Rudner$^{1}$}
\affiliation{$^1$Center for Quantum Devices, Niels Bohr Institute, University of Copenhagen, 2100 Copenhagen, Denmark \\
$^2$Department of Theoretical Physics, University of Geneva, 1211 Geneva, Switzerland \\
$^3$Physics Department, Technion, 320003 Haifa, Israel \\
$^4$Department of Physics, University of Chicago, Chicago, IL 60637, USA\\ 
$^5$Department of Condensed Matter Physics, The Weizmann Institute of Science, Rehovot, 76100, Israel}
\date{\today}
\begin{abstract}
We uncover a new family of few-body topological phases in periodically driven  fermionic systems  in two dimensions.  
These phases, which we term correlation-induced anomalous Floquet insulators (CIAFIs), are characterized by quantized contributions to the bulk magnetization from multi-particle correlations, and are classified by a family of integer-valued topological invariants. 
The  CIAFI phases do not require many-body localization, but arise in the generic situation of   {$k$-particle localization}, where the system is localized (due to disorder) for any finite number of particles up to a maximum number, $k$. 
We moreover show that, when fully many-body localized, periodically driven  systems of interacting fermions in two dimensions are characterized by a quantized magnetization in the bulk, thus  confirming the quantization of magnetization of the anomalous Floquet insulator. 
We  demonstrate our results with numerical simulations. 
\end{abstract}
\maketitle

\label{AFIClassification:Sec}

In recent years, periodic driving has been studied as a means for realizing  topological phases of matter~\cite{Yao2007,Oka2009,Kitagawa2010,Inoue2010,Lindner2011, Kitagawa2011,Gu2011,Delplace2013,Katan2013,Usaj2014,Asboth2014,Jotzu2014,
 Carpentier2015_1,Else2016a}. 
An important result of this work has been the  discovery of a wide range of intrinsically nonequilibrium
topological phases with no equilibrium counterparts~\cite{
Jiang2011,Rudner2013,TopologicalSingularities,Roy2016_1,Roy2016_2,Keyserlingk2016_1,Keyserlingk2016_2,Keyserlingk2016_3,Potter2016,Sacha_2015,Else2016a,Khemani2016,Else2016b, Choi16DTC,MonroeDTC,Po2016,MagnetizationPaper,AFAI,Quelle2017,Harper2017,EnergyPumpPaper,Martin_2017,Liu2018,Nathan_2019}. 
These ``anomalous''  phases are characterized by  robust properties of their micromotion (i.e., the dynamics that takes place within a driving period),
 such as frequency-locked oscillations in Floquet time crystals~\cite{Sacha_2015,Khemani2016,Else2016b,MonroeDTC,Choi16DTC}, or quantized orbital magnetization density in the two-dimensional anomalous Floquet-Anderson insulator (AFAI)~\cite{Rudner2013,AFAI,MagnetizationPaper}.

\begin{figure}[t]
\center{
\includegraphics[width=0.99\columnwidth]{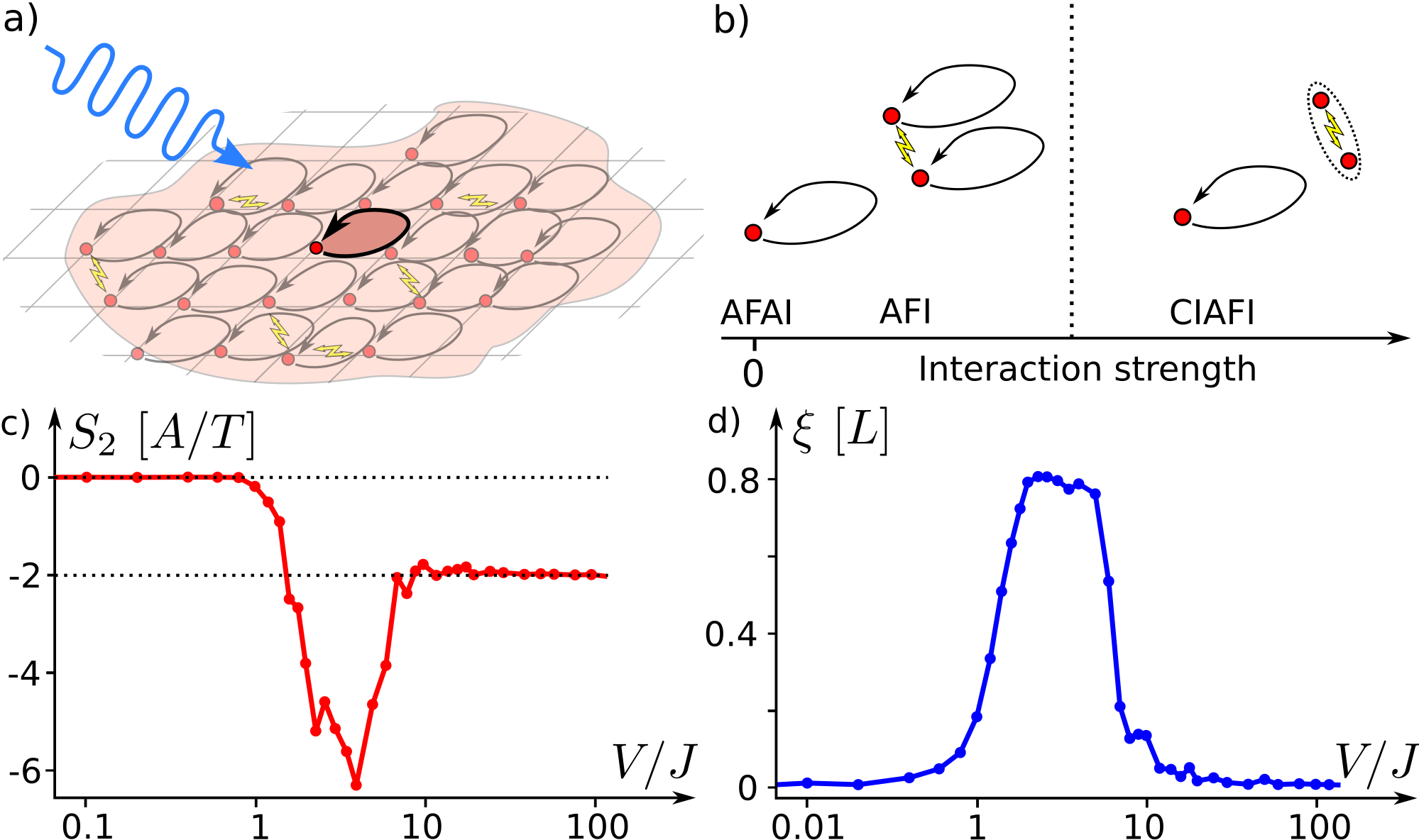}
\caption{
(a)
The anomalous Floquet insulator (AFI) is characterized by  drive-induced circulating motion of   particles in the bulk.
Nontrivial topology is revealed in  a quantized, nonzero magnetization density within regions where all states are filled, given by $\langle m\rangle = \frac{\mu_1}{T}$, where $\mu_1$ is a nonzero integer. 
(b) With sufficiently strong interactions, a new class of interaction-induced topological phases can emerge, which we term correlation-induced anomalous Floquet insulators (CIAFI's).  
CIAFI phases   are characterized by a quantized, nonzero contribution to the  magnetization from $\ell$-particle correlations.
Such correlations can for example arise due to immobilization  of many-particle bound states, as depicted in the figure. 
(c,d)
Topological phase transition between the AFI and a CIAFI phase with $\mu_2 =2$ obtained from numerical simulations of a driven Hubbard-like model (see Sec.~\ref{sec:numerics} for details). 
(c) Contribution to the time-averaged magnetization in the system due to two-particle correlations, $S_2$ (see Sec.~\ref{sec:ady_section} for definition and relationship with $\mu_2$), as a function of the interaction strength $V$. 
(d) The correlation length $\xi$ in the system diverges for interaction strength $V$ comparable to the hopping $J$, indicating a topological transition between AFI and CIAFI phases. 
}
\label{fig:Fig1}\label{fig:mu2_numerics}}
\end{figure}

Disorder plays a crucial role for stabilizing Floquet 
phases in closed systems. 
In particular, in the presence of interactions, disorder-induced many-body localization (MBL) provides a mechanism for the system to avoid uncontrollably absorbing energy  from the driving field, and thereby to retain nontrivial properties at long times~\cite{Lazarides2015,Ponte2015,Abanin20161}. 
Importantly, the requirement of many-body localization does not preclude the system from exhibiting a variety of types of symmetry-breaking and topological order~\cite{Else2016b,Khemani2016,Nathan_2019}.

In this paper we characterize the topological properties of time-evolution in  two-dimensional periodically driven systems of fermions
which exhibit either full many-body localization, or a weaker form of ``$k$-particle localization'' that we define below~\cite{Nathan_2019,Lazarides2015,Abanin20161,Ponte2015}  (see Fig.~\ref{fig:Fig1}).
Recent results  suggest that this class of systems can support a nontrivial topological phase, known as the 
 Anomalous Floquet Insulator~\cite{Nathan_2019} (AFI), which can be seen as the generalization of the AFAI to interacting systems (see Refs.~\onlinecite{AFAI,MagnetizationPaper}). 
Despite being localized and insulating, the AFI features nontrivial circulating currents in the bulk, which in the noninteracting case (the AFAI) give rise to quantized orbital magnetization~\cite{MagnetizationPaper}. 
In a geometry with boundaries,  the AFI supports thermalizing chiral edge states coexisting with a localized bulk~\cite{AFAI,Nathan_2019}.
The existence AFI as a stable many-body state of matter rests on the existence of MBL; even if MBL does hold out to infinite times, the phenomenology of the AFI is expected to persist for at least exponentially long times.

The motivation of our work is to determine the topological invariant(s) that characterize the AFI.
Focusing on the topological characterization of the micromotion of particles in the bulk (i.e., the dynamics which take place within each driving period), we uncover two main results.

As our first result, we confirm that, like the AFAI, the AFI is characterized by a quantized magnetization density in regions of the bulk where all states are occupied, as schematically depicted in Fig.~\ref{fig:Fig1}a.
Specifically, the magnetization density is quantized as ${\mu_1}/T$ where $T$ denotes the driving period, and ${\mu_1}$ is an integer characterizing the topological phase.
This quantization is protected by many-body localization, and ${\mu_1}$ cannot change under any  deformation of the system that preserves MBL.

As the second major finding of our work, we uncover a rich new structure of topological invariants that emerges in the interacting case:
while  periodically driven systems of noninteracting fermions in two dimensions (such as the AFAI) may be characterized by a single invariant ${\mu_1}$, 
their interacting counterparts  are characterized by a {\it family} of integer-valued topological invariants $\mu_1,\mu_2,\ldots$. %
The invariant $\mu_\ell$ encodes information about the contribution to the time-averaged magnetization from $\ell$-particle correlations.
Hence, interactions allow for a richer topological structure in the system. 

The topological protection of the invariant $\mu_\ell$ relies on a less restrictive notion of localization than the conventional notion of MBL.
Specifically, $\mu_\ell$ is well-defined and topologically protected when all Floquet eigenstates with up to $k$ particles are localized {for some $k\geq l$}.  
We term this notion of localization  ``$k$-particle localization.'' 
Many-body localization corresponds to $k$-particle localization in the limit where $k$  and the system size  goes to infinity, while allowing the particle density to be finite in the thermodynamic limit.
While the existence of MBL in more than one dimension is still a subject of debate~\cite{DeRoeck_2017},  $k$-particle localization for finite $k$ is well established  in any dimension~\cite{Aizenman_2009}. 
It is likely that systems exhibiting $k$-particle localization, even if not fully MBL, may still display long-lived transient phenomena: delocalization in such systems must be induced by $k+1$-particle correlated processes, whose rates are expected to be exponentially suppressed in $k$ for sufficiently weak interactions.

Our results above show that $k$-particle localized Floquet systems of interacting fermions in 2D are characterized by $k$ independent topological invariants, $\mu_1 ,\ldots\mu_k$.
When one or more of the  higher-order invariants  are nonzero, the system is in a new, strongly-correlated, intrinsically nonequilibrium phase  that is topologically distinct from  any noninteracting system, including the (noninteracting) AFAI.
 We term this class of phases Correlation-Induced AFIs (CIAFIs).
Here we consider a broader notion of the term ``phase'' than for equilibrium systems; 
in the sense we consider here, a  phase characterizes the structure of  the  Hamiltonian of the
 {\it isolated} system, independently of the  particular state of the system (and in particularly, independently of 
particle density and temperature). 


 We present a  family of models which interpolate from the  AFI phase to a CIAFI phase with a nonzero value of $\mu_2$, and  demonstrate  the existence of a nontrivial CIAFI phase in the model through  numerical simulations [see Fig.~\ref{fig:mu2_numerics}(c)-(d)].

%
The arguments leading to the identification of the higher-order invariants $\mu_\ell$ can in principle also  be applied to bosonic systems where the total number of bosons is conserved (e.g., as in systems of bosonic atoms in optical lattices).
Hence AFI and CIAFI phases also exist for $k$-particle localized bosonic systems.
However, for simplicity, in this paper, we consider fermionic systems only.

The rest of the paper is organized as follows. 
In Sec.~\ref{sec:ady_section}, we summarize the main results of this paper.
In Sec.~\ref{StructureOfFloquetOperator:sec:stab} we briefly review the structure of the Floquet operator in many-body  and $k$-particle localized  systems, and of the orbital magnetization operator. 
In Sec.~\ref{IndentificationOfInvariants:sec:stab} we use the time-averaged magnetization density operator to  identify a set of topological invariants $\{\mu_{\ell}\}$ that characterize the AFI phase, and 
show that nonzero values of the invariants give rise to a quantized magnetization density in regions where all sites are occupied (Sec.~\ref{MagnetizationInFullyOccupiedRegions:sec}). 
In Sec.~\ref{sec:numerics} we present a family of models that realize both the AFI and CIAFI phases, and support  our conclusions with numerical simulations of these models.
We conclude with a discussion in Sec.~\ref{Discussion:sec}.

\section{Summary of main results}
\label{sec:ady_section}
We begin by summarizing the main results of this paper.
We consider a two-dimensional periodically driven systems of interacting fermions, which is $k$-particle (or many-body) localized due to disorder~\cite{fn:accidental_resonances}.
To characterize the topology of the system, we quantify the circulating motion of particles in the bulk. 
This circulating motion can be captured through the time-averaged magnetization density operator of each plaquette $p$ in the Heisenberg picture,  $\bar m_p$. 
The magnetization density $\bar m_p$ measures the total time-averaged current that circulates around the plaquette; see Sec.~\ref{sec:floquet_operator} for a definition of this operator and  a review of its properties. 
From its intrinsic properties,  we show that the  trace of $\bar m_p$   defines a family of topological invariants for the system.
Specifically, the trace of $\bar m_p$ in the $\ell$-particle subspace, $\Tr_\ell\, \bar m_p$, for each $\ell = 1,\ldots k$, must take the same value for each plaquette in the system;  this value  cannot change under any smooth deformation of the parameters of the system that preserves $k$-particle localization.
Hence   $\Tr_\ell \,\bar m_p$ for each $\ell = 1,\ldots k  $ constitutes a topological invariant of the system. 
The intrinsic invariants $\mu_1\ldots \mu_k$ described  in the introduction   are constructed {by forming system-size independent, integer-valued combinations of} the (system size dependent) invariants $\Tr_1 \, \bar m_p, \ldots  \Tr_k\, \bar m_p$;   see Sec.~\ref{sec:cumulants} for further details.

To illustrate the physical meaning of the  invariants $\{\mu_\ell\}$,  consider first the case where the system  holds a single fermion, initially located on site $i$ in the lattice (we assume, without loss of generality, that each site holds a single orbital). When all single-particle Floquet eigenstates  are localized,  the particle will remain confined near site $i$ at all times. 
However,  the driving field may cause the particle to undergo circulating motion, as schematically depicted in the bottom left of Fig.~\ref{fig:Fig1}(b). 
This circulating  motion gives rise to a nonzero long-time-averaged (orbital) moment, $\bar M_i$. 
For    both single- and many-particle systems (which we consider below), the total time-averaged magnetic moment can be computed as the integral of magnetization density over the entire lattice, $\sum_{p} \bar m_p a^2$. 
Ref.~\cite{AFAI} showed that the sum of $\bar M_i$ over all single-particle states, $S_1 \equiv \sum_{i} \bar M_i$, is quantized as an integer times $A/T$, where $A$ denotes the area of the system; this integer defines $\mu_1$.
As an implication, magnetization density is quantized in the bulk of the system in regions where all states are occupied.

We now consider the dynamics resulting from initializing the system in a two-particle state where   sites $i$ and $j$ are occupied.
We let $\bar M_{ij}$ denote  the total long-time-averaged magnetization of the  system resulting from this initialization.
In the absence of interactions, one can verify that $\bar M_{ij}  = \bar M_i +\bar  M_j$. 
However,  with interactions  present,
$\bar M_{ij}$  generically differs from $\bar M_i+\bar M_j$ when sites $i$ and $j$ are close to each other. 
The deviation can be measured by the  ``magnetization cumulant'' $C_{ij} \equiv \bar M_{ij}-(\bar M_i+\bar M_j)$. 
In Sec.~\ref{IndentificationOfInvariants:sec:stab} below,  we show that, when all $1$- and $2$-particle states are localized,  the sum of $C_{ij}$ over all distinct two-particle configurations, $S_2 \equiv \sum_{i< j} C_{ij}$, must be {\it quantized}, as an integer $\mu_2$ times $A/T$.
The number $\mu_2$  cannot change under any perturbation that preserves localization of  states with $1$ and $2$ particles.
Thus,   $\mu_2$  is a topological invariant protected by $2$-particle localization, and characterizes the {contribution to the} magnetization associated with $2$-particle correlations. 
The higher-order invariants, $\mu_\ell$ for $\ell >2$, are defined analogously to $\mu_2$ from higher-order ``cumulants'' of the magnetization  (see Sec.~\ref{sec:cumulants} for details), and $\mu_\ell$ is protected under any perturbation that preserves $\ell$-particle localization.

We term the class of phases characterized by nonzero values of the higher-order invariants (i.e., $\mu_\ell$ for $\ell >1$) as correlation-induced anomalous Floquet insulators (CIAFIs).
The AFI phase is the MBL extension of the noninteracting AFAI, where all higher-order invariants must be zero, and  can thus only be characterized by a nonzero value of $\mu_1$. 
Hence the CIAFI phases are distinct from the AFI. 

In Sec.~\ref{sec:numerics} we present  a model that realizes a CIAFI phase with $\mu_{2}=-2$. 
The model consists of spin-$1/2$ fermions on a bipartite square lattice {with Hubbard-like on-site interactions and disorder}, subject to the $5$-step driving protocol of the canonical AFAI model~\cite{Rudner2013,AFAI,MagnetizationPaper} [see Fig.~\ref{fig:numerics}(a)]. 

As discussed in Sec.~\ref{sec:numerics}, and shown numerically in Fig.~\ref{fig:Fig1}(c), the strength of the Hubbard-type interactions, $V$, controls the topological phase of the model [see Fig.~\ref{fig:Fig1}(b)]:
when interactions are absent ($V=0$), the system is in the  AFAI phase with $\mu_1 =2$, while  all higher-order invariants take value zero~\cite{AFAI}. 
When interactions are weak, but finite, our numerical results  indicate that many-body localization  persists, and hence the system remains in the AFI phase with $\mu_1 = 2$ (here the factor of $2$ accounts for the two spin species). 
In particular, the values of all higher-order invariants must remain zero [$S_2 = 0$, see Fig.~\ref{fig:Fig1}(c)]. %
However, when  interactions are much stronger than the tunneling rate between the sites, $J$, they act to block tunneling to or from doubly-occupied sites, resulting in nonzero values of $C_{ij}$ for such configurations.
We demonstrate that this effect drives the model into a CIAFI phase with $\mu_2=-2$ ($S_2 =  - 2 A/T$).
In Fig.~\ref{fig:Fig1}(d), we confirm that the transition between the AFI and CIAFI phases in this model is accompanied by a divergence of the localization length of the two-particle states of the system.

\section{Many-body and $k$-particle localization in periodically driven systems}
\label{StructureOfFloquetOperator:sec:stab}
\label{sec:floquet_operator}
The main result of this work is to characterize the topological properties of time-evolution in two-dimensional periodically-driven $k$-particle (or many-body) localized fermionic systems. 
%
As a preliminary step, in this section we 
review the  structure of the Floquet operator in such systems. 

The system we study is a  two-dimensional lattice systems of interacting fermions, of physical dimensions $L\times L$, subject to periodic driving. 
While our results apply to any type of lattice, below we  assume for simplicity that the system is defined on a square lattice with lattice constant $a$ and (time-dependent) nearest-neighbor tunneling.
The time evolution of the system is described by the time-periodic Hamiltonian $H(t)=H(t+T)$, where $T$ is the driving period. 
To avoid complications from the coexistence of thermalizing chiral edge states and a  localized bulk~\cite{Nathan_2019}, we focus on the case where the system is defined on a torus, such that no edges are present~\cite{fn:bulk_edge_coexistence}.

\subsection{Structure of Floquet operator in many-body localized systems}
\label{sec:liom_structure}
We first review  the structure of the Floquet operator when the system is  many-body localized, i.e., when any state of the system exhibits localized behavior in the thermodynamic limit. 
The concepts we introduce here also  form a basis for our discussion of the more general case of $k$-particle localization (Sec.~\ref{sec:k_particle_localization}).

When the system is MBL, it has a complete set of  emergent  local  integrals of motion~\cite{Serbyn2013,Huse2014,Ponte2015,Abanin20161} (LIOMs), $\{\hat n_a\}$.
 The LIOMs form a mutually commuting set of  quasilocal operators that are individually preserved by the stroboscopic evolution of the system~\cite{fn:intermediate_time_localization}.
The number of independent LIOMs in the localized system is given by the dimension $D_1$ of the  system's single-particle Hilbert space. 
For spinless fermions with one orbital per site, we have $D_1= L^2/a^2$.  
The LIOMs $\{\hat n_\alpha \}$ may thus be labelled by a single index $\alpha $ which runs from $1$ to $D_1$.

To make the discussion more concrete, the LIOMs can be identified from the system's Floquet operator~\cite{Ponte2015}, $U(T)$.
The Floquet operator is defined as the  evolution operator of the system, $U(t)\equiv \mathcal T e^{-i \int_0^t\! dt \, H(t)}$, {evaluated for a time interval corresponding to one complete} 
  driving period $T$.
 Here $\mathcal T$ denotes the time-ordering operation, and we work in units where $\hbar = 1$ throughout.
Analogously to nondriven systems,  the stroboscopic time-evolution (i.e., the time-evolution at integer multiples of the driving period $T$)
  is conveniently expressed in terms of the eigenstates  of the Floquet operator, $\{|\psi_n\rangle\}$, known as {\it Floquet eigenstates}. 
These satisfy $U(T)|\psi_n\rangle = e^{-i\varepsilon _n T}$, where $\varepsilon _n$ has units of energy and is known as {quasienergy}. 
Note that each  quasienergy $\varepsilon _n$  is only defined modulo the driving frequency $\Omega \equiv 2\pi/T$. 
The stroboscopic time-evolution is hence equivalent to that generated by the static {effective Hamiltonian}, $H_{\rm eff}\equiv \sum_n \varepsilon _n|\psi_n\rangle\langle \psi_n|$, since $U(T)=  e^{-i H_{\rm eff} T}$.

In the many-body localized regime,  the effective Hamiltonian  takes the form
\begin{align}
H_{\rm eff}=\sum_{\alpha_1} \varepsilon_{\alpha_1} \hat n_{\alpha_1} +\sum_{\alpha_1,\alpha_2}\varepsilon_{\alpha_1\alpha_2}\hat n_{\alpha_1} \hat n_{\alpha_2} +\cdots \ .
\label{eq:FloquetOperatorForm}
\end{align}
Each coefficient  $\varepsilon_{\alpha_1 \ldots a _\ell}$ (referred to as a quasienergy coefficient in the following) is  associated with a particular combination  $ \n $ 
formed from the $D$ distinct LIOMs, and has  units of energy. 
Each sum $\sum_\al$ in Eq.~\eqref{eq:FloquetOperatorForm} runs over all $\binom{D}{\ell}$  combinations of $\ell$ distinct {LIOMs}, 
  where $\binom{a}{b}$ denotes the binomial coefficient.
 The above form of the Floquet operator implies that each LIOM $\hat n_\alpha $ is preserved by the stroboscopic evolution of the system, and thus the operators $\{\hat n_\alpha \}$ are integrals of motion.

We now review some important properties of the LIOMs which we  use in the following. 
Firstly, each  LIOM $ \nhat_\alpha $  can  be written in the form of a fermionic counting operator: $\hat n_\alpha = \hat f^{\dagger}_\alpha \hat  f_\alpha $.
Here $\hat f_\alpha $ is a  (dressed) quasilocal fermionic annihilation operator, constructed from the original lattice annihilation and creation operators $\{\hat c_i\}$ and $\{\hat c^\dagger_i\}$, respectively, as:
$\hat f_{\alpha } =  \sum_{i}\psi ^{ \alpha }_{ i }\hat  c_i  + \sum_{i jk} \psi^\alpha _{ijk } \hat c^\dagger_i  \hat c_j  \hat c_k   + \sum_{i \ldots m} \psi^{\alpha}_{ijklm}\hat c^\dagger_i\hat  c^\dagger_j\hat  c_k \hat c_l \hat c_m+ \cdots$,
where $\hat c_i $ annihilates a fermion on site $i $ in the lattice.   
Through the identification of the LIOMs with fermionic counting operators, we note that $\sum_\alpha  \hat n_\alpha $ gives the total number of fermions in the system.

Another crucial property of the LIOMs is that each LIOM $\hat n_\alpha $ has its support localized around a particular location $\vec r_\alpha $ in the lattice. 
Specifically,  the magnitude of the coefficient $\psi^{\alpha}_{i_1 \ldots i_\ell}$  decreases exponentially with the distance $s$  from any of the sites $i _1 ,\ldots i _\ell$ to $\vec r_\alpha $: $\psi^{\alpha }_{i_1 \ldots i_\ell} \sim e^{-s /\xi_f}$, where  the length scale $\xi_f$ sets the spatial extent of the LIOMs.
Similarly to the LIOMs, the quasienergy coefficients $\{\varepsilon _{\al}\}$ also exhibit  localized behavior. 
Specifically, $\varepsilon _\al$ decays as $e^{- d /\xi_\varepsilon }$, where $d$ is the distance between any two of the LIOM centers $\vec r_{\alpha_1} \ldots \vec r_{\alpha_k}$; here $\xi_\varepsilon $ is another localization length scale (not necessarily identical to $\xi_f$, see Ref.~\onlinecite{Abanin2018}).

As is evident above, MBL systems may be characterized by several distinct localization lengths~\cite{Abanin2018}.
In particular, the LIOM expansion above establishes two length scales, $\xi_f$ and $\xi_\varepsilon $. 
In the following, we will make use of an additional relevant length scale, $\xi_l$, which characterizes the spread of time-evolved operators.

\subsection{$k$-particle localization}
\label{sec:k_particle_localization}
As we explained in the introduction, the topological classification we develop in this work applies to a more general class of  systems than those exhibiting full MBL; 
specifically, the invariants we identify 
 can be defined   for any system that is $k$-particle localized for some nonzero $k$. 
As defined in the introduction,  $k$-particle localization is understood as the situation where all Floquet eigenstates holding 
$\ell$ particles for $\ell =1,\ldots k$ are localized. 
In the remainder of this paper we will make use of similar notation, such that $\ell$ always refers to a specific particle-number sector, while $k$ refers to the ``degree of localization'' of the system: i.e., $k$ is  defined as the integer such that Floquet eigenstates in the system with $k$ or fewer particles are localized, while at least one Floquet eigenstate with $k+1$ particles is delocalized.


For $k$-particle localized systems, 
we expect a LIOM decomposition and effective Hamiltonian $H_{\rm eff}$ as defined in Eq.~\eqref{eq:FloquetOperatorForm} can be written to describe the evolution in Fock space of up to $k$ particles, with the expansion truncated to $k$th order.
Full MBL can be seen as a special case of $k$-particle localization; specifically, MBL can be understood as the  $k \to \infty$ limit of $k$-particle localization   where the localization length of the truncated LIOM expansion described above remains bounded for all $k$.

\section{Topological invariants of the time evolution}
\label{sec:theory}
\label{IndentificationOfInvariants:sec:stab}
In this section, as the main result of our work, we characterize the micromotion of $k$-particle  localized systems {(which includes the case of MBL as described above)}.
We show that such systems may exhibit non-trivial micromotion, featuring steady-state circulating currents at long times.
We characterize these circulating currents by analyzing the time-averaged magnetization density operator of the system.
From this analysis we identify a set of topological invariants $\mu_{1}\ldots \mu_k$ that characterize the steady-state circulating currents that the system may support. 

In a stepwise fashion, below we  consider the dynamics of a $k$-particle localized system in the $\ell$-particle subspace for each $\ell = 1,\ldots k$ (allowing $k$ to be infinite for fully MBL systems). 
This approach ensures that  our our results do not rely on full MBL to be valid, 
 while still applying to this class of systems if such exist.

\subsection{Characterization of micromotion} 

\label{sec:micromotion}

\begin{figure}
	\includegraphics[width=\columnwidth]{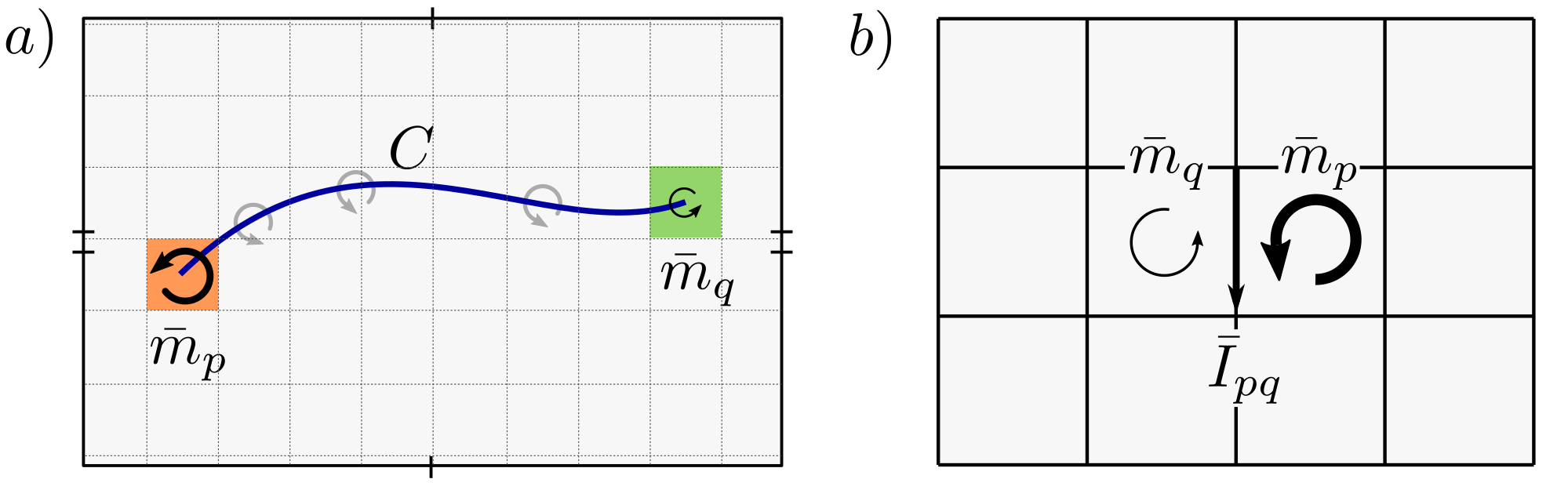}
	\caption{
	a) Schematic depiction of the relationship between current and magnetization density [Eq.~\eqref{eq:IcResult}]. 
	In many-body localized systems,  the time-averaged current passing through a cut $C$  is determined by the difference between the currents circulating around the cut's two end-points, $p$ and $q$. 
	The currents circulating around plaquette $p$ are measured by the magnetization density operator $\bar m_p$. 
        b) Ampere's law on the lattice. 
The difference in magnetization densities between two adjacent plaquettes $p$ and $q$ gives the current $\bar I_{pq}$ on the bond between them.
} 
	\label{fig:MagnetizationDensityFig}
\end{figure}

To characterize the micromotion of $k$-particle localized  systems, in this subsection we consider the dynamics within the subspace of states holding $\ell$ particles, where $\ell \leq k$.
Naively,  one might expect that the time-averaged current density in this subspace 
   always vanishes due to localization. 
Indeed, 
 there can be no net flow of charge across any closed curve.  
 However, for an {\it open} curve (or ``cut''), as schematically depicted in Fig.~\ref{fig:MagnetizationDensityFig}a, a nonzero time-averaged current may run across the cut due to uncompensated local circulating currents around the curve's endpoints.
The total current circulating around a point in a given plaquette is precisely
the magnetization density in this plaquette.

To establish this relationship in more rigorous terms, we consider the total time-averaged current that passes through a cut $C$ between plaquettes $p$ and $q$ in the lattice, as  depicted in Fig.~\ref{fig:MagnetizationDensityFig}a.
The operator $I_C(t)$ measuring  the current through the cut $C$ is given by 
\be 
I_C(t) = \sum_{b \in B_C} I_{b}(t),
\label{eq:CutCurrentDef}
\ee
where $I_{b}$ denotes the bond current operator on bond $b$ (restricted to the $\ell$-particle subspace)~\cite{fn:bond_orientation}, and the sum runs over the set $B_C$ of all bonds that cross the cut $C$ [see Appendix \ref{app:IcForm} for an explicit definition of $I_b(t)$].
Note that $I_{b}(t)$, and thereby $I_C(t)$, depends on time in the Schr\"odinger picture due to the explicit time-dependence of the Hamiltonian $H(t)$.

To characterize the circulating currents in the system, we seek the long-time-averaged expectation  value of the current $\llangle I_C\rrangle$ for an arbitrary
initial $\ell$-particle state, $|\psi\rangle$. 
Here we introduce the notation $\llangle \mathcal{O}\rrangle \equiv \lim_{\tau \to \infty} \frac{1}{ \tau} \int_0^{\tau} dt\, \langle \psi(t)|\mathcal{O}(t) |\psi(t)\rangle$ to indicate the time-averaged expectation value in the  state $|\psi(t)\rangle$.  
The time-averaged current $\llangle I_C\rrangle$ may equivalently be computed in the Heisenberg picture  as $\llangle I _C \rrangle = \langle \psi| \bar I_C |\psi\rangle$, where  $|\psi\rangle$ denotes the initial many-body state of the system, and $\bar I_C$ denotes the long-time-average of the current operator $I_C$ in the Heisenberg picture:
\be 
\bar {\ I_C}  = \lim_{\tau \to \infty} \frac{1}{\tau} \int_0^\tau\! \!\! {\rm d}t\,  U^\dagger(t)   I_C(t) U(t),
\label{eq:OperatorTimeAverageDef}
\ee 
where $U(t)$ denotes the system's time-evolution operator as defined above. 
For later, we define $\overline {\mathcal O } \equiv  \lim_{\tau \to \infty} \frac{1}{\tau} \int_0^\tau \!\! dt\,  U^\dagger(t)  \mathcal O(t) U(t)$ for any operator $\mathcal{O}$.

As argued above, the time-averaged current $\bar I_C$ across cut $C$ can only have nonzero expectation value   due to localized circulating currents at the cut's two endpoints, $p$ and $q$. 
This implies that $\bar I_C$   only depends on the details of the system near plaquettes $p$ and $q$. 
In Appendix \ref{app:IcForm} we verify this intuition,   by  proving that the operator $\bar I_C$ only has support near the  two endpoints of the cut $C$.
Specifically, assuming only $k$-particle localization and  conservation of charge, we show that, within the $\ell$-particle subspace, where $\ell\leq k$, $\bar I_C$   must take  the form
\be 
\bar I _C = \bar m_p - \bar m_q , 
\label{eq:IcResult}
\ee
where the operator $\bar m_p$ has its full support  (up to an exponentially small correction) within a distance $\xi_l$ from plaquette $p$, and similarly for $\bar{m}_q$.
Here   $\xi_l$ is a finite, system-size independent length scale  measuring the spread of operators in the system (within the $\ell$-particle subspace): 
specifically,  for any time-periodic operator $A(t)$ with a finite region  of support $R$, the long-time average $\bar A$ (when restricted to the $\ell$-particle subspace) is a local integral of motion  with support within a finite distance $\xi_l$ from $R$ (up to an exponentially small correction)~\cite{FN:ConstructionOfLIOMs}.

Crucially, the operator $\bar m_p$ in Eq.~\eqref{eq:IcResult}   is the same for {\it any} cut with an endpoint in plaquette $p$. 
Thus, Eq.~\eqref{eq:IcResult} uniquely defines the  operator $\bar m_p$  for each plaquette $p$ in the system, up to a correction exponentially small in system size. 
Specifically, let plaquette $q$ be separated from plaquette $p$ by a distance $d$, of order the system size, $L$. 
In this case,  $\bar m_p$ can be identified uniquely from the terms of $\bar I_C$ which have support nearest to plaquette $p$, up to a correction of order $\mathcal O(e^{-d/\xi_l})\sim \mathcal O(e^{-L/\xi_l})$.

For each plaquette $p$, $\bar m_p$  may be defined from Eq.~\eqref{eq:IcResult} as described above by considering a cut of length $\sim L$ (up to an exponentially small correction).
The set of operators $\{\bar m_p\}$ obtained in this way then obey Eq.~\eqref{eq:IcResult} for {\it any} two plaquettes in the lattice. 
 In particular, when the plaquettes $p$ and $q$ are adjacent, Eq.~(\ref{eq:IcResult}) 
  implies that $ \bar m_p  -  \bar m_q  = \bar I_{pq}$, where $ \bar I_{pq}$ measures the time-averaged current on the bond separating plaquettes $p$ and $q$, as schematically depicted in Fig.~\ref{fig:MagnetizationDensityFig}b.
This relationship is the time-averaged lattice version of  Ampere's law, which  relates the  current density, $\vec j$, to the magnetization density, $\vec m$: $\vec j = \nabla \times \vec m$ (see Ref.~\onlinecite{MagnetizationPaper}).
We thus identify the operator $\bar m_p$  as the time-averaged magnetization density in the system at plaquette $p$~\cite{NoteHowever}.
As the above discussion shows, the time-averaged magnetization $\bar m_p$ measures the total current circulating around plaquette $p$.

\subsection{Topological invariance of $\Tr_k\, \bar m_p$}
\label{InvarianceOfmk:sec}

%
We now show that, for each value of $\ell=1,\ldots k$, the trace of $\bar m_p$ in the $\ell$-particle subspace, $\Tr_\ell \bar m_p$, takes the same value for all plaquettes in the system. 
Subsequently (in Sec.~\ref{ValuesOfTheInvariants:sec}) we show that this universal value is quantized as an integer multiple of  $1/T$, $z_\ell$. 
Periodically driven $k$-particle localized systems of fermions in two dimensions are thus characterized by the $k$ integer-valued topological invariants  
$z_1 \ldots z_k$.

We prove the topological invariance of $\Tr_\ell\, \bar m_p$ through a simple line of arguments.
First, Eq.~\eqref{eq:IcResult} implies:
\be 
\Tr_\ell\, \bar m_p - \Tr_\ell\, \bar m_q = \Tr _\ell\, \bar I_C. 
\ee
Using the cyclic property of the trace and $U(t)U^\dagger(t) = {\bf 1}$, we find $\Tr_\ell\, \bar I_C = \lim_{\tau \to \infty} \frac{1}{\tau} \int_0^\tau \!\!dt\, \Tr_\ell\, I_{C}(t) $. 
Recall from Eq.~\eqref{eq:CutCurrentDef} that the current operator $I_C(t)$ is given by a sum of bond current operators. 
Noting that any bond current operator $I_b(t)$  is by construction traceless (see Appendix \ref{app:IcForm}), we conclude that $\Tr_\ell\, \bar I_{C}=0$. 
Hence we find:
\be 
\Tr_\ell\, \bar m_p  = \Tr_\ell\, \bar m_q. 
\label{barmpCorrespondence:eq:stab}\ee
This relation  holds for {\it any} pair of plaquettes in the lattice. 
Therefore, {for a given disorder realization}, $ \Tr_\ell\, m_p$ must take the same  universal value for all plaquettes in the system.

We now show that the universal value of $\Tr_\ell\, \bar m_p$ is a topological invariant of the system 
 in the thermodynamic limit ($L \to \infty$)~\cite{FN:ExponentiallySmallChangeOfM}.
Consider perturbing $H(t)$ within some subregion $R$ of the system (by a small but finite amount),  in such a way that $\ell$-particle localization is preserved. 
 Before and after the perturbation, $\Tr_\ell\,\bar m_p$  only depends on the details of the system around the plaquette $p$, up to an exponentially small correction (due to the exponentially decaying tails of the LIOMs).
Hence, for a plaquette $p$ located a distance of order $ L/2$ from the region $R$, 
$\Tr_\ell\, \bar m_{p_0}$  may only change by an amount of order $e^{-L/ 2\xi_l}$ due to the perturbation. 
Since $\Tr_\ell\, \bar m_{p}$ is given by the same value for {\it all} plaquettes in the system,
  $\Tr_\ell \, \bar m_{p}$  must remain unaffected by the perturbation 
even for plaquettes  within the region where  the system is perturbed, $R$. 
Thus,  $\Tr_\ell \, \bar m_p$ 
 is unaffected by any local perturbation that preserves $\ell$-particle localization, up to a correction  exponentially suppressed in system size. 
We  conclude that $\Tr_\ell\, \bar m_p$ is a topological invariant of the system, protected by $\ell$-particle localization.

In the following, it is convenient to parameterize the topologically-invariant value of  $\Tr_\ell \,\bar m_p$ by a dimensionless number; we hence let $z_\ell$ denote the value of  $\Tr_\ell\, \bar m_p$  in units of the inverse driving period,  such that $\Tr_\ell \, \bar m_p = z_\ell/T$. 

\subsubsection{Quantization of $z_\ell$}
\label{ValuesOfTheInvariants:sec}
Here we show that the dimensionless invariant 
$z_\ell$ must take an {integer value for each $\ell$}.
To do this, 
 we use an approach that generalizes the one employed for the noninteracting case in Ref.~\onlinecite{MagnetizationPaper}. 
This subsection provides a 
 summary of the  proof, while full details  are given in Appendix~\ref{app:stab:MagneticFluxResponse}.

To begin, we consider the total time-averaged magnetization operator, $\bar M \equiv \sum_p \bar m_p a^2$.
Since $\Tr_\ell\, \bar m_p $ takes the 
 value  $z_\ell/T$ for all plaquettes in the system, we have 
\be
\Tr_\ell \bar M =z_\ell\,  L^2  /T.
\label{eq:bar_m_z_ell_correspondence}
\ee 
To  establish the quantization of $z_\ell$, we  proceed in two steps.
First, we obtain  $\Tr_\ell \bar M $  
 from the response of  the system to the insertion of the weak uniform magnetic field $B_0 = 2\pi/L^2$ that corresponds to one flux quantum piercing the torus (note that the flux quantum is given by $2\pi$ in the units we employ): we show that, in the thermodynamic limit,
\be 
 e^{-i\Tr_\ell ( \bar M) B_0 T} = |\tilde U(T)|_\ell / |U(T)|_\ell,
 \label{eq:mag_trace_relation}
\ee
where $\tilde U(T)$ denotes the Floquet operator of the system in the presence of the  magnetic field $B_0$, and $|\,\cdot\,|_\ell$ denotes the determinant within the $\ell$-particle subspace. 
Subsequently, we show that the determinants $|\tU|_\ell$ and $|U|_\ell$ must be identical (see also Ref.~\onlinecite{MagnetizationPaper}); this implies that    $\Tr_\ell (\bar M) B_0 T $ equals an integer multiple of $2\pi$. 
Using $B_0 = 2\pi/L^2$ along with Eq.~\eqref{eq:bar_m_z_ell_correspondence}, we  conclude that  $z_\ell$ must be an integer.

To obtain Eq.~\eqref{eq:mag_trace_relation} (which forms the first step in our derivation), we show that  the magnetic moment of each $\ell$-particle Floquet eigenstate, $|\psi_n\rangle$, 
gives the response of its quasienergy, $\varepsilon _n$, to the addition of the weak magnetic field $B_0$. 
Letting $\tilde \varepsilon _n$ denote the  perturbed quasienergy level in the one-flux system associated with $|\psi_n\rangle$
(see the following for  details, and, in particular, for a discussion of the perturbation-induced  resonances), we show in Appendix~\ref{app:stab:MagneticFluxResponse}  that
\be
\tilde \varepsilon _n  - \varepsilon _n \approx -  \langle \psi_n|  \bar M|\psi_n\rangle B_0 .
\label{eq:qe_approx_relation}
\ee
Specifically, the sum of $\tilde \varepsilon _n - \varepsilon _n$ over {\it all} $\ell$-particle Floquet states satisfies
\be 
\sum_n(\tilde \varepsilon _n - \varepsilon _n)  = - \sum_n \langle \psi_n|  \bar M|\psi_n\rangle B_0 + \mathcal O ( e^{-L/\xi}), \label{eq:qe_response}
\ee
where $\mathcal O ( e^{-L/\xi})$ denotes some (dimensionfull) correction which goes to zero as $e^{-L/\xi}$ in the thermodynamic limit.  
We obtain Eq.~\eqref{eq:mag_trace_relation}  from Eq.~\eqref{eq:qe_response}   by multiplying with $-iT$, taking the exponentials on both sides and recalling  that $|\tilde U(T)|_\ell = \exp(-i\sum_n \tilde \varepsilon _n T)$ and likewise for $U(T)$.

Eq.~\eqref{eq:qe_response} can be obtained through first-order perturbation theory in $B_0$.
In Appendix~\ref{app:stab:MagneticFluxResponse}, we provide a rigorous derivation  of this result,  along with an exact definition of the one-to-one relationship between the quasienergy levels of the one- and zero-flux systems which   Eq.~\eqref{eq:qe_response} implicitly requires. (In particular, we give the prescription for uniquely identifying $\tilde \varepsilon _n$ for each ``unperturbed'' quasienergy level $\varepsilon _n$.).
Here we  summarize the arguments: near the  region of support  of $|\psi_n\rangle$~\cite{fn:region_of_support},  the Hamiltonian of the one-flux system, $\tilde H(t)$, is given by $H(t)  - \sum_{b}{I_{b}(t)}\theta_b + \mathcal O(\theta_b^2)$, where $\theta_{b}$ denotes the Peierls phase  on bond $b$ induced by the magnetic field $B_0$,  and $I_b(t)$ denotes the bond current operator (see Sec.~\ref{sec:micromotion} and Appendix~\ref{app:IcForm}). 
 Note that there is a gauge freedom in choosing the Peierls phases; we choose them to be of order $1/L^2$ near the region of support of $|\psi_n\rangle$ (such that the subleading correction in the above expansion of $\tilde H(t)$ can be neglected in the thermodynamic limit).

In the thermodynamic limit $L\to \infty$, one may naively
 expect that 
the quasienergy spectrum of the one-flux system can be obtained through a first-order perturbative expansion in $\delta H(t) \equiv \tilde H(t) - H(t)$  (for each $|\psi_n\rangle$), which is approximately identical to $-\sum_{b}{I_{b}(t)}\theta_b $. 
However, note that the 
 convergence of such an expansion  to first order is only ensured if the ratio  between the matrix elements of $\delta H$ in the  Floquet eigenstate basis and the corresponding quasienergy level spacings, 
$r_{mn}\equiv \langle \psi_m|\delta H(t)|\psi_n\rangle/(\varepsilon _m-\varepsilon _n)$, is much smaller than $1$ for {\it all} choices of $\ell$-particle Floquet eigenstates $m$ and $n$. 
While the perturbation $\delta H(t)$ is of order $L^{-2}$, the many-body level spacing in the $\ell$-particle subspace   is of order $\Omega/(L^{2\ell})$, where $\Omega \equiv 2\pi/T$ denotes the angular driving frequency.
Hence, in the thermodynamic limit $r_{mn}$ can potentially be much larger than $1$ for certain choices of $m$ and $n$. 
However, in Appendix~\ref{app:stab:MagneticFluxResponse} we provide a careful analysis   that confirms our initial  expectation:  with a probability that goes to $1$ in the thermodynamic limit (for each $\ell$ between $1$ and $k$), $r_{nm}$ goes to zero for {\it all} choices of $m$ and $n$.
This result arises because  states where $\langle \psi_n|\delta H|\psi_m\rangle$ is  nonvanishing must be spatially close, and hence experience local level repulsion. 

The above discussion  shows that the quasienergy level  corresponding to the state $|\psi_n\rangle$ in the one-flux system,  $\tilde \varepsilon _n$,  is captured by 
  first-order perturbation theory with respect to $\delta H(t)$. 
Expanding the quasienergy $\tilde \varepsilon _n$ to first order in $\delta H(T)$, we obtain 
\be 
\tilde \varepsilon_n  - \varepsilon _n 
\approx \frac{1}{T} \int_0^T\!\!\!{\rm d}t\, \langle \psi_n| U^\dagger(t) \delta H(t) U(t)|\psi_n\rangle
\label{eq:first_order_expansion}
\ee
 (see Appendix~\ref{app:stab:MagneticFluxResponse} for proof).
Using $\delta H(t) \approx  -\sum_b \theta _b I_b(t)$ along with the fact that in a Floquet eigenstate the time-averaged expectation value over one period is identical to the long-time average, we find 
\be 
\tilde \varepsilon_n  - \varepsilon _n  \approx  - \sum_b \theta_b \langle \psi_n|\bar I_b|\psi_n\rangle, 
\label{eq:deltaepsilon}
\ee
where $\bar I_b$ denotes the long-time average of the bond current $I_b(t)$ in the Heisenberg picture (see Sec.~\ref{sec:micromotion}).

Recall from Eq.~\eqref{eq:IcResult} (see also Fig.~\ref{fig:MagnetizationDensityFig}b) that $\bar I _b = \bar m_{p_b} - \bar m_{q_b}$, where $p_b$ and $q_b$ denotes the two adjacent plaquettes separated by the bond $b$, such that $b$ is oriented counterclockwise with respect to $p_b$~\cite{fn:bond_orientation}.
Inserting this result into Eq.~\eqref{eq:deltaepsilon}, we note that each plaquette in the lattice appears four times exactly (namely once for each of the four bonds bounding the plaquette). 
Rearranging the terms from a sum over bonds to a sum over plaquettes, we thus find 
\be 
\tilde \varepsilon_n  - \varepsilon _n \approx  - \sum_p \langle \psi_n| \bar m_p|\psi_n\rangle  (\theta_{b_{p,1}}+\theta_{b_{p,2}}+\theta_{b_{p,3}}+\theta_{b_{p,4}}).
\label{eq:magdensres}
\ee
where $b_{p,i}$ denotes the lattice bond that constitutes the $i$th edge of plaquette $p$ (counted in clockwise order starting from the positive $x$-direction), and   $\theta_{b_{p_i}}$ gives the Peierls phase acquired by traversing the bond counterclockwise with respect to $p$. 
The sum of Peierls phases $\theta_{b_{p,1}}+\theta_{b_{p,2}}+\theta_{b_{p,3}}+\theta_{b_{p,4}}$ hence gives the flux through plaquette $p$, and hence yields exactly $B_0 a^2$ for each plaquette.
Eq.~\eqref{eq:qe_approx_relation} follows by using    $\bar M \equiv\sum_p a^2 \bar m_p  $.

The rigorous derivation in Appendix~\ref{app:stab:MagneticFluxResponse} shows that the correction to the approximate equality in Eq.~\eqref{eq:qe_approx_relation} scales with system size as $L^{-4}$, and hence is subleading in thermodynamic limit (recall that $B_0 \sim L^{2}$).
We subsequently use the  LIOM structure of the Floquet operator in Eq.~\eqref{eq:FloquetOperatorForm} to show that, remarkably, these individual corrections  approximately cancel out when summed over all $\ell$-particle states, yielding an {\it exponentially} suppressed net correction, which scales with system size as $e^{-L/\xi}$. 
This establishes Eq.~\eqref{eq:qe_response}, and thereby also Eq.~\eqref{eq:mag_trace_relation}. 

What remains to be shown is that $U(T)$ and $\tilde U(T)$ have identical determinants in the $\ell$-particle subspace.  
We show this using the approach from Ref.~\onlinecite{MagnetizationPaper}:  the determinant of any time-evolution operator can be found from the time-integrated trace of the Hamiltonian~\cite{TopologicalSingularities}: $|U(T)|_\ell=e^{ -i \int_0^T\!\! dt'\,  \Tr _\ell H(t)}$.
This follows because
\be 
\sum_n \varepsilon _n = -\frac{i}{T} \int_0^T \!\!{\rm d}t\, \Tr[U^\dagger(t) \partial _t U(t)]_\ell,
\ee
which can be straightforwardly verified using  the spectral decomposition of $U(t)$.
Identifying the integrand in the right-hand side above as $-i \Tr[H(t)]_\ell$, we find  $|U(T)|_\ell=\exp( -i\int_0^T \!\!{\rm d}t\, \Tr[H(t)]_\ell)$.
Since the insertion of a magnetic flux only modifies  off-diagonal elements of the Hamiltonian (in the  lattice site basis), the trace of the Hamiltonian is unaffected by the magnetic field $B_0$.
Thus $|\tU(T)|_\ell = |U(T)|_\ell$. 
Hence, the right-hand side of Eq.~\eqref{eq:mag_trace_relation}  equals $1$ and therefore the argument in the exponent of $e^{-i {\rm Tr}_\ell (\bar{M}) B_0 T}$ must be an integer multiple of $2\pi$. 
Combining this with Eq.~\eqref{eq:bar_m_z_ell_correspondence} and using that $B_0 = 2\pi/L^2$, we conclude that $z_\ell$ must be an integer.
\subsection{Cumulant basis of invariants}
\label{sec:cumulants}
The above discussion shows that $k$-particle localized systems are characterized by the $k$ independent, integer-valued topological invariants $z_1\ldots z_k$.
Here $z_\ell$ gives the trace of the magnetization density operator in the $\ell$-particle subspace (in units of the inverse driving period).
However, each $z_\ell$ depends on the size of the system, 
and thus is  not an {\it intrinsic} property of the system. 
 For instance, in noninteracting systems, $z_\ell$ scales as $L^{2(\ell-1)}$, where $L$ is the physical dimension of the system~\cite{FN:ThisFollowsFromEqs1718}.
In this subsection we construct linear combinations of the invariants $z_1\ldots z_k$ that give an equivalent set of system size {\it independent} invariants $\mu_1\ldots \mu_k$  that characterize the {\it intrinsic} topological properties of the system. 

The intrinsic invariants $\mu_1\ldots \mu_k$ can be expressed as the cumulants of the magnetization operator, as discussed in Sec.~\ref{sec:ady_section}. 
To illustrate, consider the time-averaged magnetic moment, $\bar M \equiv \sum_p a^2 \bar m_p$, of a state where two particles are initialized on sites $i$ and $j$, which we denote $\bar{M}_{ij}$. 
The average of the total magnetic moment, taken over {\it all} $2$-particle states, is given by {$\frac{1}{D_{2}}(z_2L^2/T)$}, where $D_\ell$ denotes the dimension of the $\ell$-particle subspace. 
For each $i$ and $j$, we write $\bar M_{ij} = \bar M_i + \bar M_j + C_{ij}$, where, as in Sec.~\ref{sec:ady_section}, $\bar M_i$ denotes the time-averaged magnetization of the system holding a single particle initially located at site $i$.
From this definition of $C_{ij}$, we find 
\be 
\frac{1}{L^2} \sum_{i<j} C_{ij} =   \frac{z_2 - 2(L^2-1)z_1}{T},
\label{eq:second_order_cumulant}
\ee
where we used that  $\Tr_\ell \bar M = z_\ell L^2/T$ for $\ell = 1,2$.
The right hand side is evidently an integer multiple of $1/T$. 
We take this integer to be our definition of the intrinsic invariant $\mu_2$.

Note that $\mu_2$ gives the mean value of $S_i \equiv \sum_{j\neq i} C_{ij}$ over all sites $i$ (recall that $C_{ij}=C_{ji}$).
  Importantly, due to the fact that the two particles only influence each other's motion when they are within a localization length of one another, the cumulant $C_{ij}$  is only significant for $\mathcal O(\xi_l^2/a^2)$ choices of $j$ for each $i$.
    The mean value of $S_i$ is therefore an intrinsic quantity, which does not depend on the system size; in particular, it remains finite in the thermodynamic limit. 
In the noninteracting case, $C_{ij} =0 $, and $\mu_2 = 0$. 
Thus, $\mu_2$ gives the contribution to the magnetization from $2$-particle correlations.

We extend this definition to higher numbers of particles, by expanding $\bar M$ in terms of the fermionic annihilation and creation operators, $\{\hat c_i\},\{\hat c_i^\dagger\}$. 
Since $\bar M$ preserves the number of particles, we have 
\be 
\bar M = \sum_{i_1j_1} \mathcal M_{i_1;j_1} \hat c^\dagger_{i_1}\hat  c_{j_1} +\!\!\! \sum_{i_1i_2,j_1j_2} \!\!\! \mathcal M_{i_1i_2;j_1j_2}\hat c^\dagger_{i_1} \hat c^\dagger_{i_2} \hat c_{j_1}\hat c_{j_2}  + \cdots. 
\label{eq:bar_m_expansion}
\ee
{Without loss of generality, we take $\mathcal M_{i_1\ldots i_k;j_1\ldots j_k}$ to be nonzero only if $i_1 < i_2 \ldots <  i_k$ and $j_1 >j_2 \ldots> j_k$, such that each independent combination of creation and annihilation operators appears only once in the above sum.
 We see that  the expectation value of $\bar M$ in a single-particle state $|i\rangle \equiv \hat c^\dagger_i |0\rangle$ (where $|0\rangle$ denotes the vacuum state) is given by $\mathcal M_{i;i}$. 
We thus  identify $\mathcal M_{ii} = \bar M_i$, where $\bar M_i$ was defined above. 
Likewise, in the two-particle-state $|ij\rangle \equiv\hat  c^\dagger_i\hat  c^\dagger_j |0\rangle$ (where $i<j$), the expectation value of $\bar M$ is given by $\mathcal M_{i;i}+\mathcal M_{j;j} + \mathcal M_{ij;ji}$. 
We thus identify  $\mathcal M_{ij;ji} = C_{ij}$.
The higher-order cumulants can be defined in a similar fashion, such that $C_{i_1 , \ldots i_\ell} = \mathcal M_{i_1\ldots i_\ell;i_\ell \ldots i_1}$.
Note that the long-time average of an operator in the Heisenberg picture, such as $\bar M$, must be diagonal in the Floquet eigenstate basis; for example, $\mathcal M_{i;j}$ is diagonal in the basis of single-particle Floquet eigenstates.}

Due to localization and the locality of interactions (see above), the coefficient $C_{i_1 \ldots i_\ell}$ can only be nonzero if all sites $i_1\ldots i_\ell$ are spatially close (on the scale of  $\xi_l$). 
Thus, through arguments analogous to those below Eq.~\eqref{eq:second_order_cumulant}, for each $\ell = 1\ldots k$, $\frac{T}{L^2}\sum_{i_1,\ldots i_\ell}C_{i_1 \ldots i_\ell}$ is a (dimensionless) intrinsic quantity of the system. 
This motivates us to define the $\ell$-th intrinsic invariant as: 
\be 
\mu_\ell = \frac{T}{L^2} \sum_{i_1 \ldots i_\ell} C_{i_1 \ldots i_\ell}.
\ee

To relate $\mu_\ell$ to the invariants $z_1\ldots z_k$, we take the $\ell$-particle  trace in Eq.~\eqref{eq:bar_m_expansion}.
Using  $\Tr_\ell [\hat c^\dagger_{i_1}\ldots\hat c^\dagger_{i_\nu}\hat  c_{i_\nu}\ldots\hat  c_{i_1} ] =\binom{D_1-\nu}{\ell- \nu} $ (this can be verified from combinatorial arguments), where $D_1 = L^2$ denotes the dimension of the system's single-particle subspace, we find 
\be 
z_\ell=\sum_{\nu=1}^\ell \binom{D_1-\nu}{\ell-\nu}  \mu_\nu ,
\label{TrkMucorrespondence:eq:stab}
\ee
where we used  $\Tr_\ell \bar M = z_\ell L^2 /T$.
By induction, one can  verify that each $\mu_\ell$ is an integer.
First, by the definition above, $\mu_1 $ equals $z_1$, and hence is an integer.
For $\ell >1$, $\mu_\ell = z_\ell - \sum_{\nu=1}^{\ell-1}\binom{D_1-\nu}{\ell-\nu}  \mu_\nu$. 
Thus, if $\mu_1\ldots \mu_{\ell-1}$ are integers, $\mu_\ell$ is also an integer (since $z_\ell$ is  an integer).

To further elucidate the physical meaning of the intrinsic invariant 
$\mu_\ell$, we express it in terms of the LIOMs that were introduced in Sec.~\ref{sec:floquet_operator}.
Since the long-time average of any Heisenberg picture operator is diagonal in the basis of Floquet eigenstates~\cite{FN:InCaseOfDegeneracies}, the operator
$\bar m_p$ must be an integral of motion~\cite{Chandran2015}.
This requires  $\bar m_p$ to take the following form   in terms of the of the LIOMs $\{\hat n_\alpha \}$ that we introduced in Eq.~\eqref{eq:FloquetOperatorForm}: 
\be 
\bar m_p =   \sum_{\alpha_1} m^{p}_{\alpha_1}\hat  n_{\alpha_1} + \sum_{\alpha_1 \alpha_2} m^p_{\alpha_1 \alpha_2} \hat n_{\alpha_1} \hat n_{\alpha_2} + \cdots\ .
\label{BarMpExpansion:eq:stab}
\ee 
Here, for each term involving a products of $\ell$ LIOMs, 
 the sum $\sum_{\alpha_1 \ldots \alpha_\ell}$ runs over the $\binom{D_1}{\ell}$ distinct combinations  of  $\ell$ LIOM indices $\alpha_1 \ldots \alpha_\ell$. 
Due to the finite support of the operator $\bar m_p$, we note  that the coefficient $m^{p}_{\alpha_1 \ldots \alpha_\ell}$
 vanishes as $e^{-d/\xi_l}$, where $d$ is  the distance from the plaquette $p$ to the center of the most remote of the LIOMs $\alpha_1 \ldots \alpha_\ell$.

Taking the $\ell$-particle trace in Eq.~\eqref{BarMpExpansion:eq:stab} and using $\Tr_\ell [\hat n_{\alpha _1}\ldots \hat n_{\alpha _\nu}] = \binom{D-\nu}{\ell -\nu}$, we find 
\be 
z_\ell=\sum_{\nu=1}^\ell \binom{D_1-\nu}{\ell-\nu} \sum_{\alpha_1 \ldots \alpha_\nu} m^{p}_{\alpha_1 \ldots \alpha_\nu}/T. 
\ee
Comparing with Eq.~\eqref{TrkMucorrespondence:eq:stab} for each $\ell = 1\ldots k$, we find
\be 
\mu_\ell \equiv  \sum_{\alpha_1 \ldots \alpha_\ell} m^{p}_{\alpha_1 \ldots \alpha_\ell}/T. 
\label{MuCorrespondence:eq:stab}
\ee
Note that $\mu_\ell$ is independent of the choice of plaquette $p$.

From the expression above, it is evident that $\mu_\ell$  characterizes the intrinsic topological properties of the system.
Since the magnetization coefficients $\{m^p_{\alpha_1\ldots\alpha_\ell}\}$ vanish when the distance from any of the LIOM centers $\vec r_{\alpha_1} \ldots \vec r_{\alpha_\ell}$ to plaquette $p$ becomes large, the right-hand side of Eq.~\eqref{MuCorrespondence:eq:stab} is independent of system size in the thermodynamic limit. 
In essence, $\mu_\ell$ captures the contribution of $\ell$-body correlations to the magnetization density.

\subsection{Quantized magnetization density in fully occupied regions}
\label{MagnetizationInFullyOccupiedRegions:sec}
As a final part of this section, we show that the values of the invariants $\mu_1\ldots \mu_k$ can be measured directly from the  magnetization density within a region of the system where all sites are occupied.
In particular,  for the AFI (which is fully MBL and for which only   $\mu_1$ takes nonzero value),  the magnetization density  is  given by $\mu_1/T$.

Consider preparing the system in an $\ell$-particle state  $|\Psi_{\mathcal{R}}\rangle$  (where $\ell\leq k$) by filling all sites in some finite region of the lattice, $\mathcal{R}$, of linear dimension $d$, with all sites outside $\mathcal R$  remaining empty (here we assume this requires fewer than $k$ particles). 
For a plaquette $p$ located deep within the fully occupied region, we find the time-averaged magnetization density as $\llangle m_p\rrangle = \langle\bar m_p\rangle_{\mathcal R}$, where we introduced the shorthand $\langle \mathcal O \rangle_{\mathcal R}\equiv \langle \Psi_{\mathcal{R}}|\mathcal O|\Psi_{\mathcal{R}}\rangle$. 
Using  the  expansion of $\bar m_p$ in Eq.~\eqref{BarMpExpansion:eq:stab}, we thus find: 
\be 
\llangle  m_p\rrangle  =    \sum_{\alpha_1} {m^{p}_{\alpha_1}}\langle \hat  n_{\alpha_1}\rangle _{\mathcal R} + \sum_{\alpha_1 \alpha_2} {m^p_{\alpha_1 \alpha_2}} \langle \hat n_{\alpha_1} \hat n_{\alpha_2}\rangle_{\mathcal R} + \cdots.
\label{BarMpExpansion1:eq:stab}
\ee
To analyze the sum, we note that, for a LIOM $\hat n_a$  whose  center $\vec r_a$ 
 is located  deep within the filled region $\mathcal{R}$, all sites where  $\hat n_a$ has its support are occupied. 
 Thus
 $\hat n_\alpha |\Psi_{\mathcal{R}}\rangle = |\Psi_{\mathcal{R}}\rangle+\mathcal O (e^{-d/\xi_l})$~\cite{FN:ToSeeThis}. 
Here the correction arises from the exponentially decaying tail of $\hat n_\alpha $ outside the filled region. 
For terms in the above equation  where the centers of all the LIOMs $\alpha_1  \ldots \alpha_\nu$  are located near the plaquette $p$, the above result implies that $\langle \hat n_{\alpha_1}\ldots \hat n_{\alpha_\nu}\rangle_{\mathcal R} = 1+\mathcal O (e^{-d/\xi_l})$, since {\it all} of the LIOMs $\hat  n_{\alpha_1}\ldots \hat n_{\alpha_\nu}$  are located deep within the initially occupied region. 
For all remaining terms in Eq.~\eqref{BarMpExpansion1:eq:stab}, one or more LIOMs $\alpha_1  \ldots \alpha_\nu$ are located outside the filled region, and thus reside at least a distance $\sim d$ from the plaquette $p$.
In this case, 
 the  coefficient  $m^{p}_{\alpha_1 \ldots \alpha_\nu}$ is  exponentially small in $d/\xi_l$ [see the discussion below Eq.~\eqref{BarMpExpansion:eq:stab}]. 
For both categories of terms we can thus set $\langle \Psi_{\mathcal{R}}|m_{\alpha_1 \ldots \alpha_\nu}^p \hat n_{\alpha_1}\ldots \hat n_{\alpha_\nu} |\Psi_{\mathcal{R}}\rangle = m_{\alpha_1 \ldots \alpha_\nu}^p$, at the cost of a correction of order  $e^{-d/\xi_l}$. 
Doing so, we obtain 
\be 
\llangle  m_p\rrangle  =    \sum_{\alpha_1} m^{p}_{\alpha_1}   +  \sum_{\alpha_1 \alpha_2 } m^p_{\alpha_1 \alpha_2 }   + \ldots  +\mathcal O (e^{-d/\xi_l}).
\label{eq:LongTimeMagCalculation1}
\notag
\ee
Using Eq.~\eqref{MuCorrespondence:eq:stab}, we identify the $\ell$-th sum above as the invariant $\mu_\ell/T$.
Recalling that $\langle \Psi_{\mathcal{R}}|\bar m_p|\Psi_{\mathcal{R}}\rangle = \llangle m_p\rrangle$, we thus find:
\be 
\llangle m_p\rrangle  =   \frac{1}{T}  \sum_{\nu=1}^\ell \mu_\nu + \mathcal O (e^{-d/\xi_l}). 
\label{eq:LongTimeMagCalculation2}
\ee
The above discussion thus shows that the magnetization density deep within the filled region is given by the (convergent~\cite{FN:ToSeeThatTheSum}) sum of the invariants $\{\mu_\ell\}$.
In particular, for the AFI, where only $\mu_1$ is nonzero, $\llangle m_p\rrangle=\mu_1/T$.

We note that the individual invariants $\mu_1\ldots \mu_k$ may be extracted from the dependence of the magnetization density on the particle density in the system.  
Specifically, for a random initial state with a uniform, finite particle density $\rho$, the expectation value $\langle \hat{n}_{\alpha _1} \ldots \hat{n}_{\alpha_\nu}\rangle $, averaged over all choices of LIOMs, is given by $\rho^\nu$.
Hence, at finite particle density $\rho$, the average magnetization density in the system is given by $\llangle m_p\rrangle \approx \frac{1}{T} \sum_{\nu=1}^\ell \mu_\nu \rho^\nu$. 
The values of the individual invariants $\mu_\nu$ can thus be extracted from a fit of $\llangle m_p\rrangle$ as a function of $\rho$. 

\section{Specific model and Numerical simulations}
\label{sec:numerics}
\begin{figure}[t!]
\includegraphics[width=1\columnwidth]{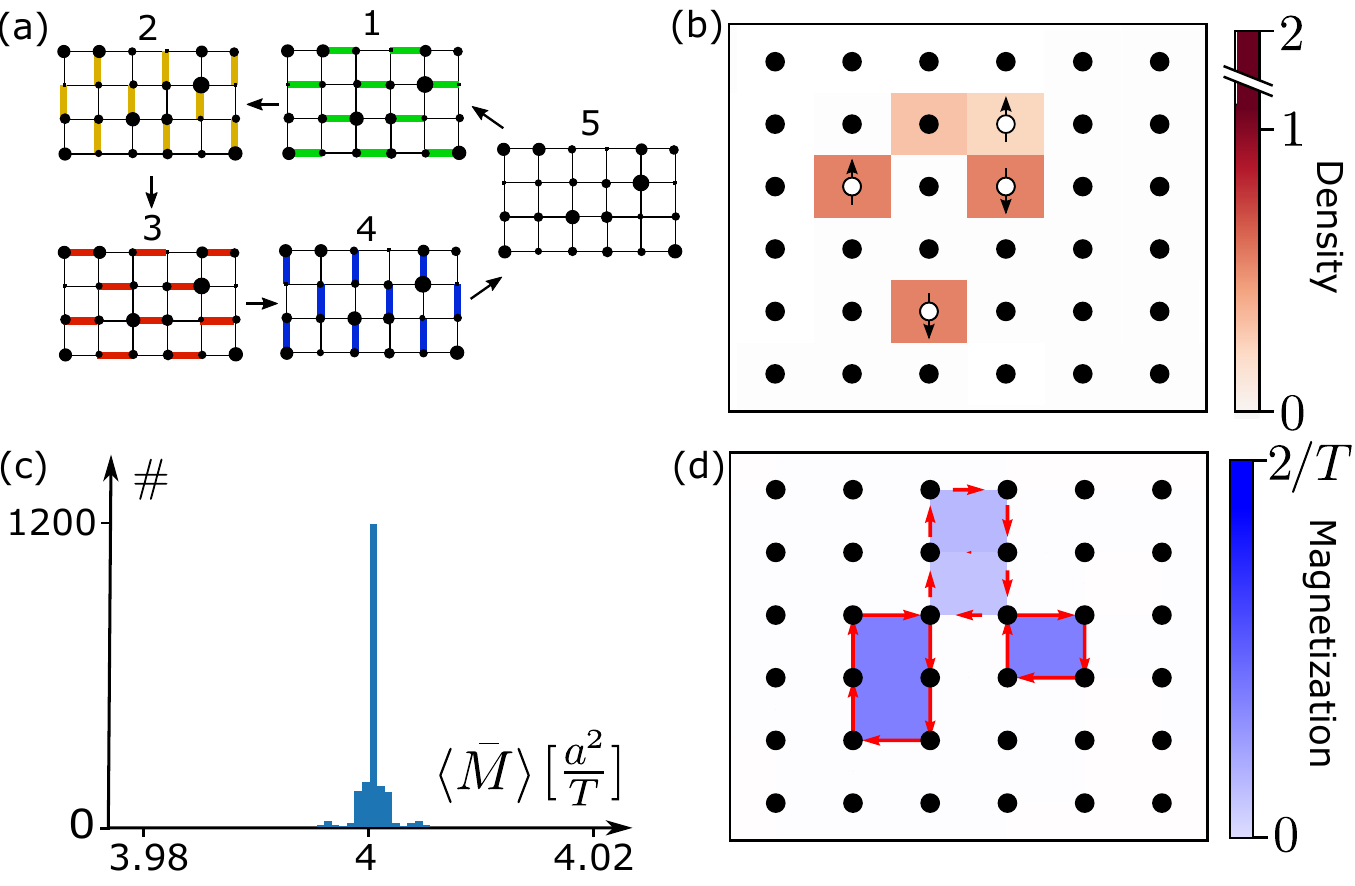}
\caption{
Simulations of model studied in Sec.~\ref{sec:numerics} in the case of weak interactions, where it realizes the AFI phase (see main text for details).  
(a) Schematic depiction of the driving protocol.
(b) Final particle density after  $1000$ driving periods for initialization in a random on-site configuration of particles (initial configuration of particles  marked by white). 
(c) Histogram of magnetic moments of $246$ randomly chosen initial states that are evolved for $1000$ periods (see main text for details). 
A single outlier at value $3.89$ is not shown here. 
(d) Time-averaged bond-current (red) and magnetization density in the system (blue) for the realization depicted in panel (a). }
\label{fig:numerics}
\end{figure}

In this section we present a simple model for a periodically driven system of interacting fermions in two dimensions, which realizes either the AFI or  a CIAFI phase. 
The model was briefly discussed in Sec.~\ref{sec:ady_section}. 
We first consider the limit of weak interaction.
In this regime we  argue that the system 
%
 realizes the AFI phase with $\mu_1 = 2$. 
Subsequently, we show that, in the limit of strong interactions,
 the model is characterized by a quantized, nonzero value of the ``two-particle cumulant'' of the magnetization density, consistently with a CIAFI phase characterized by $\mu_2 = -2$.
To support our conclusions, we provide  numerical simulations of the  model in the to above regimes. 

The model we consider consists of  fermions with spin-$1/2$ living in a two-dimensional bipartite square lattice  with periodic boundary conditions. 
The Hamiltonian   is given by 
\be
H(t) = H_{\rm dr}(t) +H_{\rm dis}+ H_{\rm int}\label{eq:model_def},
\ee
 where $H_{\rm dr}(t)$ describes piecewise-constant, time-dependent hopping,  $H_{\rm dis}$ denotes a disorder potential, while $H_{\rm int}$ describes an on-site interaction between the fermions. 
The driving protocol, which is contained in $H_{\rm dr}(t)$,  is divided into five segments, as  depicted in Fig.~\ref{fig:MagnetizationDensityFig}a.
The first four  segments each have duration $\eta T/4$, while the fifth segment has duration $(1-\eta ) T$;
the parameter $\eta $ is a number between $0$ and  $1$  which controls the localization properties of the model (see below).
In the first four segments, $H_{\rm dr}(t)$ turns hopping on for the four different bond types in a counterclockwise fashion,  as indicated in Fig.~\ref{fig:numerics}a, while $H_{\rm dr}(t) = 0$ in the fifth segment. 
More specifically, in the 
$j$-th segment  (where $j\leq 4$),
\be \label{eq:Hid}
H_{\rm dr}(t) = J\sum_{ \vec r \in A}\sum_{s} (\hat c^\dagger_{\vec r + \vec b_j,s}\hat c_{\vec r,s} + {\rm h.c.}).
\ee
Here $\hat  c_{\vec r,s}$ annihilates a fermion on site $\vec r$ with spin $s$, and the vectors $\{\vec b_j\}$ are given by 
 $\vec b_1=-\vec b_3= (a,0)$ and $\vec b_2 =- \vec b_4 = (0,a)$. 
The $\vec r$-sum above runs over all sites in  sublattice A of the bipartite square lattice.
We set the tunneling strength to $J = \frac{2\pi}{\eta  T}$, such that, in the absence of disorder and interactions, $H_{\rm dr}$ would generate a perfect transfer of particles across the active bonds in each of the first four segments.
The parameter $\eta $ controls how rapidly the ``hopping $\pi$-pulses'' are applied (and thereby how strong they are relative to the disorder and interaction potentials), and thus controls the localization properties of the model;  smaller $\eta$ yields stronger localization (see Ref.~\onlinecite{Nathan_2019}).

The  disorder and interaction terms $H_{\rm dis}$ and $H_{\rm int}$  are constant throughout the driving period and are given by 
\begin{equation}\label{eq:Hdis0}
H_{\rm dis}=\sum_{{\bf r},s} w_{\bf r} \hat\rho_{\vec r,s}, \quad H_{\rm int}=V\sum_{{\bf r} } \hat \rho_{\bf r,\uparrow}\hat \rho_{{\bf r,\downarrow}}.
\end{equation}
For each site, $w_{\bf r}$ takes a random value in the interval $  [-W, W]$, and $\hat\rho_{\vec r,s}\equiv \hat c^\dagger_{\vec r,s} \hat c_{\vec r,s}$ denotes the occupancy on site $\vec r$.
The parameter $V$ has units of energy and denotes the strength of the interactions. 
Note that when $V\gg J$, tunneling is effectively blocked between doubly-occupied and vacant sites.
As we show below, this blocking leads to a nonzero value of the higher-order invariant $\mu_2$. 

To characterize the topological properties of the model, we  consider  the dynamics of particles in the two limits of weak and strong interactions. 
Below we demonstrate how these two regimes drive  the model  into the AFI phase with $\mu_2 =2$ and a CIAFI phase with $\mu_2= -2$, respectively.
We substantiate these conclusions with numerical simulations in Sec.~\ref{sec:simulations}.

In the absence of interactions, $V= 0$,
the model in Eq.~\eqref{eq:model_def} reduces to two decoupled copies of the AFAI model from Ref.~\onlinecite{AFAI}.
When interactions are weak, but nonzero, Ref.~\onlinecite{Nathan_2019} suggests that the phase remains MBL (i.e., non-thermalizing). 
Since the model should be connected to the non-interacting AFAI, we hence expect the system to be in the AFI phase~\cite{Nathan_2019} with winding number $\mu_1 = 2$ 
(see also discussion in Sec.~\ref{sec:ady_section}). 
The factor of $2$ arises from the extra species of fermions introduced due to the spin-$1/2$ degree of freedom.

We now show that the  model above is in a CIAFI phase with $\mu_2= -2$ in the limit of strong interactions, $V\to \infty$.
To see this,  we consider  the time-averaged magnetic moment $\bar M_{ij}$     (see Sec.~\ref{sec:cumulants}) that results
  when initially occupying  two single-particle states $i$ and $j$, where each choice of $i$ or $j$ corresponds to  a particular site and spin. 
Recall that tunneling is blocked  when the first particle is located on, or tunnels to, a site occupied by the second particle. 
Hence, doublons (i.e., states where two particles occupy the same site) remain frozen in place, implying that $\bar M_{ij}=0$   if  $i$ and $j$ correspond to  the same site being occupied. 
For all other initial  configurations, interactions effectively
do not affect the dynamics, and one can verify that $\bar M_{ij}  =\bar M_{i} + \bar M_{j}$, where $\bar M_i$ denotes the time-averaged magnetic moment in the single-particle state $i$.
As a result, the ``cumulant'' $C_{ij}\equiv \bar M_{ij} -\bar M_i - \bar M_j$ takes value $-2a^2/T$ when the 
 initialization $ij$ corresponds to a doublon configuration, and value zero for all other $2$-particle initializations (see Sec.~\ref{sec:cumulants} for definition of $C_{ij}$).
We recall from Sec.~\ref{sec:cumulants} that $\mu_2 = S_2 T/L^2$, where $S_2 \equiv \sum_{i<j} C_{ij} T/L^2$. 
Since there are $L^2/a^2$ distinct doublon configurations, where $L$ denotes the physical dimension of the lattice, we find that $S_2 = -2 L^2/T$.
Thus, $\mu_2 =- 2$ in the limit  $W=0$, $V\to \infty$. 
From the discussion in Sec.~\ref{sec:theory}, we expect the  quantization of $\mu_2$ to persist for finite disorder, $W$, and finite (but large) values of the interaction strength, $V$.

The discussion above shows that the model in Eq.~\eqref{eq:model_def} is characterized by two distinct values of the invariant $\mu_2$ in the limits where $V=0$ and $V\to \infty$, respectively. 
Due to the robust quantization of $\mu_2$, which is protected by $2$-particle localization, we hence conclude that the system  supports two distinct topological phases that arise when  $V\ll J$ and $V\gg J$, respectively. 
The transition between the phases is separated by a critical point, $V_{\rm c}$~\cite{Aizenman_2009}:  when $V$ is increased past $V_{\rm c}$ in the thermodynamic limit, the localization length in the two-particle sector should diverge at $V=V_{\rm c}$, while $\mu_2$  changes abruptly from $0$ to  $-2$.

\subsection{Numerical simulations}
\label{sec:simulations}
Here we substantiate the discussion above through numerical simulations of the model:
we first consider the limit of  weak interactions, and  show that the (quantized) average magnetic moment per particle 
 remains unaffected by the nonzero interaction strength, as our analytical discussion predicts for an AFI phase with $\mu_1 = 2$.
Subsequently, we show that  that the model is characterized by a quantized, nonzero value of the invariant $\mu_2$, when $V$ is large, demonstrating that the system is in a CIAFI phase, distinct from the {$\mu_1=2, \mu_2 = 0$} AFI phase.

\subsubsection{Weak interactions:  AFI phase with $\mu_1= 2$}
\label{sec:afi_numerics}
We first present data from simulations of the model described above, in the limit of weak interactions. 
We consider a single disorder realization of the model with parameters $W=2\pi/T$, $V= 0.1 \, W$, and $\eta = 1/16 $. 
From  Ref.~\onlinecite{Nathan_2019}, we expect the model is many-body localized with these parameters.
Since the model is obtained by adding weak interactions to a model of the AFAI with winding number $2$ (see Refs.~\onlinecite{AFAI,MagnetizationPaper}; here the factor of $2$ arises because of  the spin degeneracy), we moreover expect   the system to be in the $\mu_1 =2$ AFI phase (i.e., with $\mu_\ell = 0$ for $\ell >1$). 

{
To probe the topology of the system, we compute the mean magnetic moments of random time-evolved $4$-particle states in a lattice of $6\times 6$ sites.
The long-time averaged magnetic moment, introduced in Sec.~\ref{sec:theory},  is defined as $\bar M = \sum_p a^2 \bar m_p$. 
 The mean expectation value of $\bar M$, averaged over 
  randomly chosen $\ell$-particle states (i.e., states chosen randomly from a given orthonormal basis) is given by $M_0[\ell] \equiv\binom{D}{\ell}^{-1}\Tr_\ell\, \bar M$, where  the binomial coefficient $\binom{D}{\ell}$ counts the number of possible $\ell$-particle states in the system of $D = 2 L^2 $ single-particle states (here the factor of $2$ arises due to the spin degeneracy, and $L=6$ for the case we consider).
Using 
that $\Tr_\ell \bar M = z_\ell L^2/T$, along with Eq.~\eqref{TrkMucorrespondence:eq:stab}, we can express  $M_0[\ell]$ in terms of the topological invariants $\mu_1 \ldots \mu_\ell$:
 $M_0[\ell]=\frac{ L^2}{T} \sum_{\nu=1}^\ell A_\nu \mu_\nu$,
where $A_\nu =    {\binom{D-\nu}{\ell-\nu}}/ \binom{D}{\ell}$. 
For $\ell=4$ particles, our expectation that $\mu_1 =  2$ while $\mu_\ell = 0$ for $\ell >1$ hence would lead to 
\be 
\label{eq:expectation}M_0 = \frac{ 4 a^2}{T},
\ee
corresponding to an average magnetic moment per particle of $a^2/T$.
This result  was previously  established for the noninteracting limit of the model (where the system is in the AFAI phase)~\cite{MagnetizationPaper}.
The discussion above hence shows that the quantized average magnetic moment per particle in the AFAI is unaffected by  interactions, as long as the system remains in the AFI phase. }

To compute $M_0$ in  the simulation,  we pick as initial states $1972$ random configurations of four particles located on individual sites.
We evolve each initialization for $5, \!000$ driving periods with a fixed  disorder realization (the same for all initial states). 
Fig.~\ref{fig:numerics}b shows the particle density in the resulting final state for one of the realizations, after evolution for $5, \!000$ periods.
White dots and arrows indicate the corresponding initial configuration of  occupied sites and spins. 
Note that the particle density remains non-uniform and confined near the initial location of the particles, consistent with many-body localization. 
We compute the time-averaged magnetic moment $\langle \bar M\rangle $ for each of the $1972$ states, using the time-averaged bond-currents. 
The $1972$ values of $\langle \bar M\rangle $ we obtained  
in this way are plotted in the histogram in Fig.~\ref{fig:numerics}c. 
Fig.~\ref{fig:numerics}d shows the time-averaged  bond currents and magnetization density in the system for the same state used in Fig.~\ref{fig:numerics}b, used to calculate the magnetization.
The distribution of $\langle \bar M\rangle $ obtained from these initializations was found to have mean $3.999997\, a^2 /T$ and standard deviation $\delta M = 0.001a^2/T$, resulting in a standard deviation of the mean at $\delta M /\sqrt{1972} \approx 0.00003 a^2/T$. 
This result is  consistent  with a $\mu_1 =2 $ AFI phase [see Eq.~(\ref{eq:expectation})]. 

\subsubsection{Strong interactions: CIAFI phase with $(\mu_1,\mu_2)=(2,-2)$}
\label{sec:strong_interactions_model}
We now demonstrate that strong interactions drive the model into a CIAFI phase with $\mu_2 = -2$. 
These data were briefly discussed in Sec.~\ref{sec:ady_section}.
Here we present them in further detail.

To show that large interaction strength drives the model into the CIAFI phase, 
we keep $W$ and $\eta$ fixed, but vary $V$.
We moreover consider a single disorder realization with $18\times 18 $ sites.
For each value of $V$ we consider, we  obtained the time-evolution over $1000$ driving periods for 
 between 179 and 324 randomly chosen initializations where the two particles were located on particular sites and had distinct spins~\cite{fn:spin_initialization_groups}.

To establish the existence of a phase transition between the AFI and CIAFI phase, we considered the localization length in the system. 
We measured this 
using the  inverse participation ratio of the density in the  final state that resulted from each of the initializations we considered, $\mathcal P \equiv (\sum_{{\vec r}}|\rho_{{\vec r}}|^2)^{-1}$, where $\rho_{\vec r} = \sum_{s=\uparrow,\downarrow} \langle \hat c^\dagger_{{\vec r},s}\hat c_{{\vec r},s}\rangle$ denotes the particle density on site ${\vec r}$ in the final state. 
When each particle is localized on a particular site, 
$\mathcal P $ takes the value $1/4$ (in the case of a doublon configuration) or $1/2$.
In contrast, $\mathcal P = L^2/4$ indicates full delocalization (corresponding to $\rho_{\vec r} = 2/L^2$ for all ${\vec r}$).
More generally, $\mathcal P$ can effectively be seen as $1/4$ times the  number of sites where the final state has support.
This motivates us to define the effective localization length of the system, $\xi_{\rm IPR}$, as the average value of $ \sqrt{4 \mathcal P a^2} $ obtained from the initializations we probed.  

In Fig.~\ref{fig:Fig1}d, we plot the above localization length of the system, $\xi_{\rm IPR}$, as a function of $V$.
As is evident in the figure, the localization length remains small for small values of $V$.
This indicates that the $\mu_1=2$  AFI phase at $V=0$ remains stable for finite values of the interaction strength, as was also suggested by the results in Sec.~\ref{sec:afi_numerics}. 
In the range between $V=J$ and $V=10J$, the localization length diverges, consistent with a phase transition. 
For $V\gtrsim 10 J$, the localization length becomes small again, indicating the system has transitioned  back into a stable phase. 
The localization length appears to remain small as $V$ goes to $\infty $; we hence expect this new phase to be the $\mu_2 = -2$ CIAFI phase.


%
%
%
%

To verify the existence of two distinct phases (namely the $\mu_1=2$ AFI and the  $\mu_1,\mu_2 = 2,-2$ CIAFI phases), we computed the sum $S_2 \equiv \sum_{i<j} C_{ij}$, where $C_{ij} = \bar M_{ij}- \bar M_i - \bar M_j$ (see Sec.~\ref{sec:cumulants} or \ref{sec:ady_section} for definition of these quantities). 
In Fig.~\ref{fig:mu2_numerics}c, we plot the value of this sum.
The data shows a clear transition between $\mu_2 = 0$ to $\mu_2 =-2$ in the range $V = J$ to $V = 10 J$, where the localization length diverges.
This further supports the existence of a  $\mu_1,\mu_2 = 2,-2$ CIAFI phase for strong interactions, which is distinct from the AFI phase. 

%
%
%
%
%

\section{Discussion}
\label{Discussion:sec}
In this work, we characterized  the topological properties of periodically driven systems of interacting fermions in two dimensions. 
We  established that the quantized magnetization of the AFAI persists in its interacting generalization, the anomalous Floquet insulator (AFIs).
As a second result, we identified a new class of intrinsically-correlated nonequilibrium phases, namely the correlation-induced anomalous Floquet insulators (CIAFIs).
The topological invariants characterizing the CIAFIs are encoded in the multi-particle correlations of the  time-averaged magnetization density.
While this work focused on driven fermionic models and their bulk topological invariants, 
 our discussion can be readily extended to bosonic systems with particle number conservation.

Importantly, the  topological protection of the CIAFIs does not require full many-body localization, but rather relies on  {\it $k$-particle localization}, where the system is localized for any finite number of particles up to a maximum number, $k$. 
The existence of $k$-particle localization is well-established~\cite{Aizenman_2009}.
Since the existence of the CIAFI does not rely on full many-body localization, we may expect the behavior described above to be manifested via experimental signatures in the prethermal dynamics of systems which eventually thermalize at long times. 
Searching for other models that give rise to nontrivial values of these invariants and characterizing the physical properties that they imply will be interesting directions for future studies.

We demonstrated that CIAFIs may be realized in a tight-binding model with Hubbard type-interactions subject to a stepwise driving protocol. 
Recently, a noninteracting version of such a model was  experimentally realized with ultracold atoms in optical lattices~\cite{Wintersperger_2020}.
The CIAFI phases  may be achieved in a similar experiment  by adding Hubbard-type interactions to the system. 
We expect this type of interactions   is natural to implement with ultracold atoms in optical lattices.
Thus, we speculate that experimental realization of  CIAFI phases is  feasible with current  experimental platforms.


At this point it is not clear whether the CIAFI phases are compatible  with MBL, i.e., if they can exist in the thermodynamic limit of  $L\to \infty$ and $k\to \infty$. (For finite $k$, localization is possible, and the physics described above is rigorously applicable.)
In particular, we expect that CIAFI phases  will exhibit dynamics  strongly dependent on the initial
state.
In the model of Sec.~\ref{sec:numerics}, initial states where some large region $\mathcal R$ is doubly occupied would support chiral edge states moving around such regions. 
If the initial state contains such ``internal edges,'' they may thermalize and serve as a weak heat bath
for the remainder of the system. 
Next, if the density of filled regions $\mathcal R$  in the system is increased, we expect that at some point thermalizing internal edges will form a connected network, destroying localization. 
In contrast, initial states without filled, connected regions are expected to be much more stable, since there are no direct thermalization processes which involve few nearby particles;
thermalization, if it occurs at all, will proceed either due to rare thermal inclusions, or due to multi-particle tunneling into, e.g., a state with ``internal edges.'' 

After the initial posting of this work, another preprint independently classified the bulk topological properties of two-dimensional MBL systems, when particle number conservation was present~\cite{Zhang_2020}. 
Interestingly the classification in Ref.~\onlinecite{Zhang_2020} did not contain the CIAFI phases, suggesting that CIAFI phases and MBL may be incompatible. 
A definite answer for this question, however, remains lacking, and will be an interesting direction for future studies. 
In any case, the features above suggests that CIAFI phases  (rigorously established for finite particle number) may provide a versatile playground for studying the interplay of weak thermalizing baths and MBL regions, which is expected to give new insights into the stability of MBL in 2d.

The topological classification we developed in the present work relied on particle number conservation. 
Chiral phases of spins and bosons without particle number conservation, which are close relatives of the AFAI (with higher-order invariants being zero, $\mu_\ell = 0$, $\ell >  2$), were considered in Ref.~\onlinecite{Po2016}. 
It was shown that, when many-body localized, such phases are characterized by a quantized topological index which describes the pumping of quantum information along the edge over one driving period. Such an index arises from the rigorous classification of anomalous local unitary operators in one-dimensional systems, developed by Gross
et al~\cite{Gross2012}. 
It will be an interesting direction of future studies to investigate whether the bulk classification of the present work can be generalized to systems where particle conservation is not present.

In the future, it will moreover be interesting to investigate how thermalization is manifested in experimentally realistic situations for the CIAFI phases, and what  the corresponding time scales are.
With $k$-particle localization present (for some large $k$), thermalization  must be driven by correlated processes involving more than $k$ particles. 
It is natural to expect that such thermalizing process will be parametrically slow, and therefore signatures of the CIAFI phases (and the AFI), such as quantization of magnetization, would be observable even if MBL is eventually destroyed. 
A systematic study of such thermalization timescales will be an interesting question for future studies, with significance beyond the context of topological phases we considered here.

%
%

{\it Acknowledgements ---}
M.R. and F.N. thank the Villum Foundation and the European Research Council (ERC) under the European Union Horizon 2020 Research and Innovation Programme (Grant Agreement No. 678862) for support.
D.A. acknowledges support by the Swiss National Science Foundation.
N.L. acknowledges support from the European Research Council (ERC) under the European Union Horizon 2020 Research and Innovation Programme (Grant Agreement No. 639172), and from the Israeli Center of Research Excellence (I-CORE) ``Circle of Light''.
M.R. and E.B. acknowledge support from CRC 183 of the Deutsche Forschungsgemeinschaft.

\bibliographystyle{apsrev}
\bibliography{AFI_clas_bibliography}

\begin{thebibliography}{68}
\bibitem[{fn:({\natexlab{c}})}]{fn:intermediate_time_localization}
\bibinfo{note}{Note that, due to the finite Lieb-Robinson velocity of the
  system, the conservation of the LIOMs by the stroboscopic evolution requires
  the system's dynamics to also be localized at intermediate times.}
\bibitem[{fn:({\natexlab{d}})}]{fn:bond_orientation}
\bibinfo{note}{Note that, with this notation, each bond $b$ is implicitly
  defined with an orientation, such that $I_b(t)$ measures the current {along}
  the bond's orientation.}
\bibitem[{Not()}]{NoteHowever}
\bibinfo{note}{Note that Ampere's law is only meaningful when the current
  density has zero divergence. The long-time-averaged magnetization density
  $\bar m_p$ in an MBL system is always well-defined, since the time-averaged
  current density always has zero divergence. Moreover, while Ampere's law only
  defines magnetization density up to a constant shift, $\bar m_p$ is uniquely
  defined by the definition in Sec.~\ref{sec:micromotion}.}
\bibitem[{FN:({\natexlab{b}})}]{FN:ExponentiallySmallChangeOfM}
\bibinfo{note}{For a finite system, the fact that $m^p_\al$ is exponentially
  insensitive to the details of the system far away from the plaquette $p$
  means that it may only change by an amount of order $e^{-L/\xi}$ when the
  system size is increased. This implies that the sum $\Sigma_\al m^p_\al$ is
  given by its value in the thermodynamic limit, up to a correction of order
  $e^{-L/\xi}$.}
\bibitem[{fn:()}]{fn:region_of_support}
\bibinfo{note}{Here the region of support is understood as the region of the
  lattice where the particle density is significant in the state
  $|\psi_n\left.\right>$. See Appendix~\ref{app:stab:MagneticFluxResponse} for
  further details.}
\bibitem[{FN:({\natexlab{a}})}]{FN:ThisFollowsFromEqs1718}
\bibinfo{note}{For noninteracting systems, $\bar m_p$ is a one-body operator,
  and $m^p_{\a} = 0$ when $k\geq 2$. Hence only $\mu_1$ may be nonzero [see
  Eq.~(\ref{MuCorrespondence:eq:stab})]. In this case,
  Eq.~(\ref{TrkMucorrespondence:eq:stab}) implies that ${\rm Tr}_k \bar m _p =
  \binom{D-1}{k-1}\mu_1$. Using $D=L^2$, we see that $\binom{D-1}{k-1}$ scales
  as $L^{2(k-1)}$ with the system size $L$.}
\bibitem[{FN:({\natexlab{b}})}]{FN:InCaseOfDegeneracies}
\bibinfo{note}{In case of degeneracies, one can always pick a basis of
  eigenstates where $\bar m_p$ is diagonal}.
\bibitem[{FN:({\natexlab{c}})}]{FN:ToSeeThis}
\bibinfo{note}{To see this, note that $\hat
  n_\alpha|\Psi_{\mathcal{R}}\left.\!\!\right> = (1-f_\alpha
  f_\alpha^\dagger)|\Psi_{\mathcal{R}}\left.\!\!\right> $. The operator
  $f_\alpha^\dagger$ is a polynomial in $\{c_\alpha\}$ and $\{c_\alpha
  ^\dagger\}$, where each term has the net effect of creating one fermion in
  the region around LIOM $a$. Since all sites near the LIOM $a$ are occupied
  for the state $|\Psi_{\mathcal{R}}\left.\!\!\right>$, $f^\dagger_\alpha
  |\Psi_{\mathcal{R}}\!\!\left.\right> = 0$, and thus $\hat
  n_\alpha|\Psi_{\mathcal{R}}\left.\!\! \right> =
  |\Psi_{\mathcal{R}}\left.\!\!\right> $.}
\bibitem[{FN:({\natexlab{d}})}]{FN:ToSeeThatTheSum}
\bibinfo{note}{To see that the sum in Eq.~(\ref{eq:LongTimeMagCalculation1})
  converges, note that the coefficient $m_{\alpha_1\ldots\alpha_\ell}$ is
  exponentially suppressed in $d/\xi$, where $d$ is the distance from any of
  the LIOM centers $\vec r_{\alpha_1 } \ldots \vec r_{\alpha_\ell}$ to the
  plaquette $p$. The number of distinct LIOMs whose centers are located within
  a radius $\xi_l$ from the plaquette $p$ is of order $\xi_l^2/a^2$, where $a$
  is the lattice constant in the system. Therefore, the coefficient
  $m_{\alpha_1\ldots\alpha_\ell}^p$ vanishes exponentially when $\ell \gg
  \xi_l^2/a^2$. Recalling that $\mu_\ell \equiv
  \Sigma_{\alpha_1\ldots\alpha_\ell} m_{\alpha_1\ldots\alpha_\ell} ^p$ must
  take integer value for each $\ell$, we thus conclude that $\mu_\ell$ equals
  zero when $\ell\gg \xi_l^2/a^2$.}
\bibitem[{fn:({\natexlab{a}})}]{fn:spin_initialization_groups}
\bibinfo{note}{More precisely, the initializations were divided into three
  classes, that contained initializations where the two particles were located
  on the same site, adjacent sites, and all other ``on-site'' initializations,
  respectively. The average magnetization was found through the sum of the
  obtained mean values of $\langle \bar M\rangle$ within each class, weighted
  according to the number of states in the class.}
\bibitem[{fn:({\natexlab{b}})}]{fn:correction_def}
\bibinfo{note}{Specifically, we require that $\lim_{L\to \infty} L^{p'}
  \mathcal O(L^{-p}) = 0$ for $p'<p$}.
\bibitem[{FN:({\natexlab{a}})}]{FN:FluxOnTorus}
\bibinfo{note}{The fact that the total flux on the torus is given by an integer
  multiple of $2\pi$ means that this requirement does not require a
  specification of the interior region of the loop.}
\bibitem[{FN:({\natexlab{b}})}]{FN:OperatorNorm}
\bibinfo{note}{Specifically, the spectral norm of an operator $M$ is defined as
  $\norm{M} \equiv {\rm sup}_{|\psi\left.\right>} \sqrt{\left<\right.\!
  \psi|M^\dagger M |\psi \!\left.\right>/\left< \psi|\psi\right>}$.}
\bibitem[{FN:({\natexlab{c}})}]{FN:TensorProuctDefinition}
\bibinfo{note}{Here the tensor product $|\psi_1\rangle_A \otimes
  |\psi_2\rangle_B$ is defined as $\hat C^\dagger_1 \hat C^\dagger_2|0\rangle$,
  where $C^\dagger_{1}$ is the unique combination of fermionic creation
  operators that creates the state $|\psi_1\rangle_A$, i.e.,
  $|\psi_1\rangle_A=C^\dagger_1 |0\rangle_A$, where $|0\rangle_A$ denotes the
  vacuum in subsystem $A$. $C^\dagger_2$ is defined in a similar fashion.}
\end{thebibliography}

\appendix

\section{Proof of Eq.~\eqref{eq:IcResult} }

\label{app:IcForm}
\begin{figure}
	\includegraphics[width=\columnwidth]{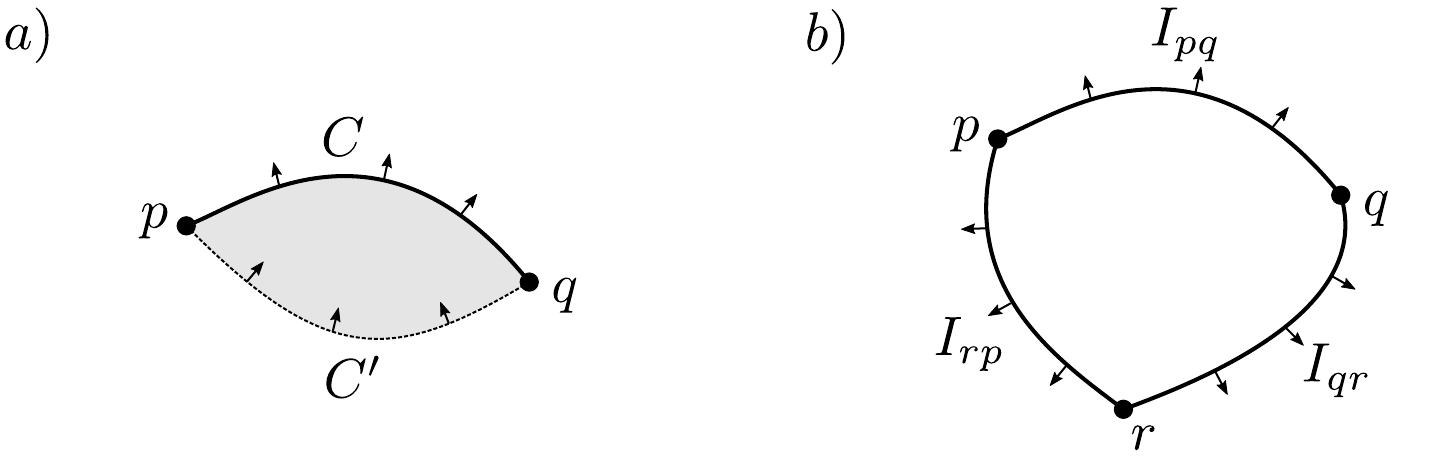}
	\caption{
a) 
Schematic depiction of the argument  
showing that time-averaged current through a cut $C$  between to plaquettes $p$ and $q$ only depends on the cut's two end-points.
Specifically, since there can be no accumulation of charge over time in the region between the cuts $C$ and $C'$, the same current must pass through the two cuts, and thus $\bar I _C  = \bar I_{C'}$ for any two cuts $C$ and $C'$ between the plaquettes $p$ and $q$.
 b) The vanishing divergence of current implies that $\bar I_{C_{pq}}+\bar I_{C_{qr}}=\bar I_{C_{pr}}$.
  }
	\label{fig:MagnetizationDensityFigApp}
\end{figure}
In this appendix we establish that the time-averaged current that passes through a cut $C$ between two plaquettes $p$ and $q$ is determined by two quasilocal operators, $\bar m_p$ and $\bar m_q$, with support centered at $p$ and $q$, respectively [see Eq.~(\ref{eq:IcResult}) and Fig.~\ref{fig:MagnetizationDensityFigApp}].
By considering two plaquettes separated by a distance much longer than the localization length, this provides a prescription for uniquely identifying the magnetization density operator $\bar{m}_p$ (up to exponentially small corrections in the distance, which can be of order the system size).

We recall from the main text that the operator corresponding to current through the cut $C$ is given by 
\be 
I_C(t) = \sum_{b \in B_C} I_{b}(t),
\label{eq:app:CutCurrentDef}
\ee
 where $I_{b}$ denotes the bond current operator on bond $b$, and the sum runs over all bonds that cross the cut $C$. 

The goal of this Appendix is to find the time-averaged expectation  value of the current, $\llangle I_C\rrangle$, resulting from some given initial state $|\psi\rangle$.
As in the main text, we use $\llangle \mathcal{O}\rrangle \equiv \lim_{\tau \to \infty} \frac{1}{ \tau} \int_0^{\tau} dt\, \langle \psi(t)|\mathcal{O}(t) |\psi(t)\rangle$.  
The time-averaged expectation value of the current $I_C$ may equivalently be computed in the Heisenberg picture 
 as $\llangle I _C \rrangle = \langle \psi| \bar I_C |\psi\rangle$, where  $|\psi\rangle$ denotes the initial state of the system.
 Here, as in the main text, 
 for any Schr\"odinger picture operator $\mathcal O(t)$ [such as $I_C(t)$], $\bar{\mathcal O}$  denotes the time-average of the current $I_C$ in the Heisenberg picture, 
\be
\bar {\mathcal O}  \equiv \lim_{\tau \to \infty} \frac{1}{\tau} \int_0^\tau \!\!\!{\rm d}t\,  U^\dagger(t)  \mathcal O(t) U(t).
\label{eq:app:OperatorTimeAverageDef}
\ee
The time-averaged current operator $\bar I_C$ is thus obtained by transforming the time-dependent operator $I_C(t)$ in Eq.~(\ref{eq:app:CutCurrentDef}) with evolution operator $U(t)$, and integrating over time as in Eq.~(\ref{eq:app:OperatorTimeAverageDef}).

To explore the properties of 
$\bar I _C$, we consider the time-averaged current for a different cut, $C'$, between the  same two plaquettes  $p$ and $q$, see Fig.~\ref{fig:MagnetizationDensityFigApp}a. 
We note that 
  $ I_C (t) -  I _{C'}(t)= \dot N_{R}(t)$, 
where $N_{R}$ measures the  number of particles in the region  $R$ between cut $C$ and $C'$ (shaded region in Fig.~\ref{fig:MagnetizationDensityFigApp}). 
Importantly, since  $N_R$ 
is bounded 
by the number of sites in the region $R$, the long-time-averaged value of  $\langle \dot N_R\rangle$ 
 must vanish. 
We thus conclude that 
$ 
 \llangle I _C \rrangle =  \llangle I _{C'}\rrangle.
$
Since this  holds for any initial state $|\psi\rangle$, we conclude that 
\be 
\bar I_C =  \bar I _{C'}.
\label{eq:TimeAveragedCurrentOperatorResult}
\ee

As a next step, we note from  Eqs.~\eqref{eq:app:CutCurrentDef}
that $\bar I_C = \sum_{b \in B_C} \bar I _{b}$, where $\bar I _{b}$  denotes the time-averaged current on bond $b$ [see Eq.~\eqref{eq:app:OperatorTimeAverageDef}].
We note that the operator $I_b(t)$ is local, with support only on the sites connected by the bond $b$.
For many-body localized systems, this implies that the operator $\bar I _{b}$ is  a localized integral of motion, with support within a  distance $\sim\xi_l$ from the bond $b$, up to an exponentially small correction~\cite{FN:ConstructionOfLIOMs}.
Hence, $\bar I_C$ is given by  a sum of terms, each of which only has support within a region of radius $\xi_l$, centered at a point along the cut $C$. 

The requirements  that  $\bar I_C$ is given by  a sum of local terms as described above,  while at the same time  taking the same value for all cuts between plaquettes $p$ and $q$ [Eq.~\eqref{eq:TimeAveragedCurrentOperatorResult}], significantly constrains the form that $\bar I_C$ can take. 
In particular, this implies that $\bar I_C = I(p,q)$, where the operator $I(p,q)$ only depends on the locations of the two plaquettes $p$ and $q$ (and not on the details of the cut $C$). 
Moreover, for any cut between plaquettes $p$ and $q$,  $I(p,q) $ is given by a sum of terms which only have support in a region of width $\xi_l$ around the cut. 
For any site located a distance larger than $\xi_l$ from both plaquettes $p$ or $q$, we can find a cut that remains separated from the site by a distance larger than $\xi_l$. 
Therefore the support of operator $I(p,q)$ can only include sites within a localization length of the endpoints $p$ and $q$.
Hence, we write:
\be 
I(p,q)= A_1(p,q)+ A_2(p,q),
\label{eq:IArelation}
\ee
 where $A_1(p,q)$ has its full support within a region of width $\xi_l$ around plaquette $p$,  and $A_2(p,q)$ has support around plaquette $q$. 
The operators $A_{1}(p,q)$ and $A_{2}(p,q)$ depend  only on the locations of plaquettes $p$ and $q$, respectively.

By letting the cut from $p$ to $q$ go through an arbitrary plaquette $r$ on the torus (see Fig.~\ref{fig:MagnetizationDensityFig}b),  we conclude from the arguments above the $I(p,r)+I(r,q)= I(p,q)$. 
This implies
\be 
A_1(p,r)+ A_2(p,r)+ A_1(r,q)+ A_2(r,q)= A_1(p,q)+ A_2(p,q).
\ee
The only terms on the left hand side with support near plaquette $r$ are the terms $A_2(p,r)$, and $A_2(r,q)$, while none of the terms on the right-hand side have support near plaquette $r$.
We thus conclude that $A_2(p,r)=-A_1(r,q)$ for any choice of two plaquettes $p$ and $q$. 
Hence we may write $A_1(r,q)=A(r)$, and $A_2(p,r)=-A(r)$ for some function $A(r)$ which only depends on the location of plaquette $r$ and has its full support near plaquette $r$. 
Using this in Eq.~\eqref{eq:IArelation}, we find
\be 
I(p,q)=  A(p)-A(q).
\ee
Identifying $A(p)= \bar m_p$, we thus conclude that Eq.~(\ref{eq:IcResult}) holds.

\section{Derivation of Eq.~\eqref{eq:qe_response}}
\label{app:stab:MagneticFluxResponse}
Here we  derive Eq.~\eqref{eq:qe_response}, 
which  is used to establish the integer quantization of the topological invariant $z_\ell$.

To recapitulate, we consider a $k$-particle localized system, where $k$ may be infinite in the case of full MBL.
For a given $\ell \leq k$, we  consider  the $\ell$-particle Floquet eigenstates of the system, $\{|\psi_n\rangle\}$, with corresponding quasienergies  $\{\varepsilon _n\}$, and let $\tilde \varepsilon _n$ denote the  perturbed quasienergy corresponding to $\varepsilon _n$  when   the weak uniform magnetic field $B_0 = 2\pi/L^2$ is inserted that results in one flux quantum piercing the torus (see below for details).
The goal of  this Appendix is to establish  two  results.
First, we show that for each $\ell$-particle Floquet eigenstate, $|\psi_n\rangle$, 
\be
\tilde \varepsilon _n = \varepsilon _n- B_0\langle \psi_n|\bar M|\psi_n\rangle+\mathcal O ( L^{-5/2}).
\label{eqa:qel_relation_aide}\ee
Here, and in the remainder of this Appendix, $\mathcal O(L^{-p})$ indicates a correction which goes to zero at least as fast as $ L^{-p}$~\cite{fn:correction_def}. 
(I.e, in the following, we only indicate  how rapidly corrections decrease  with system size.)
Secondly, we show that, when summed over all $\ell$-particle Floquet states, the corrections of order $L^{-5/2}$  in Eq.~\eqref{eqa:qel_relation_aide} approximately cancel out, yielding a net correction which is {\it exponentially} suppressed in system size:
\be 
\sum_n (\tilde \varepsilon _n - \varepsilon _n) =  - \sum_n B_0\langle \psi_n|\bar M|\psi_n\rangle + \mathcal O (e^{-L/\xi}), 
\label{eqa:qel_relation}
\ee
where $\mathcal O(e^{-L/\xi})$ likewise indicates a correction that goes to zero as $e^{-L/\xi}$ in the thermodynamic limit.

Eqs.~\eqref{eqa:qel_relation_aide}~and~\eqref{eqa:qel_relation}   implicitly require that, for each quasienergy level $\varepsilon _n$ of the (unperturbed) zero-flux system, 
it should be possible to identify a unique quasienergy level $\tilde \varepsilon _n$ of the (perturbed) one-flux system which satisfies Eq.~\eqref{eqa:qel_relation_aide}. 
In Sec.~\ref{FloquetEigenstateCorrespondence:sec:app:stab} below, we confirm that such a  complete one-to-one identification is possible  for all but a set of disorder realizations  which has measure zero in the thermodynamic limit. 

As noted in the main text, Eq.~\eqref{eqa:qel_relation} does not follow trivially from first-order perturbation theory in the weak magnetic field $B_0$: 
under a continuous perturbation of the system, the system's quasienergy spectrum undergoes exponentially many avoided crossings  due to resonances between many-body Floquet eigenstates separated by a large distance in Fock space.
Hence, first-order perturbation theory breaks down for the system.
Instead, we establish Eq.~\eqref{eq:qe_response}  with an alternative approach, using the localization properties of the many-particle Floquet eigenstates. 

In order to 
 follow this approach, we use a succession of auxiliary results which are not discussed in detail in the main text, but are crucial for the proof of Eqs.~\eqref{eqa:qel_relation_aide}~and~\eqref{eqa:qel_relation}.
  The line of arguments proceeds as follows: we  first show explicitly how the uniform magnetic field $B_0$ can be implemented in the system (Sec.~\ref{MagneticFluxImplementation:sec:app:stab}). 
Subsequently, in Sec.~\ref{ResponseOfHtoFlux:sec:app:stab} we show that, for  a given finite  region $S$ of the lattice, it is always possible to choose a gauge where the Hamiltonian $\tH$ of the  one-flux system  resembles the  Hamiltonian $H$  of the  zero-flux system  locally within $S$, and likewise for the Floquet operators $\tU$ and $U$ (Sec.~\ref{ResponseOfUToFlux:sec:app:stab}). 
Using this result, we demonstrate in Sec.~\ref{FloquetEigenstateCorrespondence:sec:app:stab} that  the  Floquet eigenstates and quasienergies, $\{|\psi_n\rangle\}$ and $\{\varepsilon _n\}$, are robust to the perturbation caused by inserting of the weak uniform magnetic field $B_0$, such that the one-to-one identification described above is possible. 
From these auxilliary results,  we prove Eq.~\eqref{eqa:qel_relation_aide} in Sec.~\ref{MagQERelationship}, and finally use Eq.~\eqref{eqa:qel_relation_aide} along with the LIOM structure of the system to establish Eq.~\eqref{eqa:qel_relation} (Sec.~\ref{seca:correction_sum}).

For the sake of brevity, throughout this Appendix we will work with a fixed degree of localization and particle number, unless otherwise noted.
Thus, in the following, $k$ and $\ell$ are   fixed constants  that  refer to the system's degree of localization and to the number of particles in the system, respectively.
We  take $\ell \leq k$ in the discussion below.

\subsection{Implementation of magnetic flux}
\label{MagneticFluxImplementation:sec:app:stab}
Here we 
discuss how the magnetic flux is implemented. 
The system we consider consists of interacting  fermions on a lattice with the geometry of a torus, of dimensions $L\times L$. 
The  Hamiltonian of the system (in the absence of a flux) 
takes the form 
\be 
H(t) = \sum_{ij }J_{ij } (t) \hat{c}^\dagger_{i }\hat c_j + H_{\rm int}(t), 
\label{eq:app:Hamiltonian:stab}
\ee
where $\hat c_i $ annihilates a fermion on site $i $ in the lattice. 
Here the first term contains both hopping and on-site potentials, including disorder, with $J_{ij}(t) = J_{ji}^*(t)$, while the  term $H_{\rm int}$ accounts for interactions. 
We allow both parts of the Hamiltonian to be time-dependent, with periodicity $T$.
To simplify the discussion, we consider the case of a square lattice model with nearest-neighbour hoppings, and a density-density interaction described by 
$H_{\rm int}=\frac12\sum_{i,j}\hat \rho_i \hat \rho_j  V_{ij }(t)$, where $\hat \rho_i  = \hat c^\dagger_i  \hat c_j  $ and $V_{ij}(t) = V_{ji}(t)$ is real.
In the  general case  of a quasilocal Hamiltonian,  the results below can also be derived using similar arguments. 

In this subsection we are interested in finding the Hamiltonian $\tH(t)$ of the system when the uniform magnetic field $B_0 = \frac{2\pi}{L^2}$ is inserted, corresponding to one flux quantum through the surface of torus. 
Having assumed $H_{\rm int}(t)$ to consist of density-density interactions, only the  first term in Eq.~\eqref{eq:app:Hamiltonian:stab}  is affected by the magnetic flux.
The Hamiltonian $\tH(t)$ thus takes the form: 
\begin{equation}
\tH (t)= \sum_{i j } e^{i \theta_{i j }} J_{i j } (t) \hat{c}^\dagger_{i }\hat c_j +  H_{\rm int}(t). 
\label{eq:JPeierlsModification}
\end{equation}
Here, the Peierls phases $\{\theta_{i j }\}$, with $\theta_{ij} = -\theta_{ji}$, must ensure that the total phase acquired by traversing a closed loop 
 on the torus is given by $B_0 A_S  \,(\mod 2\pi)$, where $A_S$ is the area enclosed by the loop~\cite{FN:FluxOnTorus}.

There are (infinitely) many distinct configurations of the phases $\{\theta_{ij}\}$ that satisfy this condition, corresponding to different choices of  gauge for the one-flux Hamiltonian $\tH(t)$. 
As the starting point for the following discussion, we consider the following  Landau-type gauge:
let $\theta_i ^x$ denote the Peierls phase for hopping along the bond in the positive $x$-direction from site $i $ (and similarly let $\theta_i ^y$ be the Peierls phase for hopping in the positive $y$-direction), and give them the values: \be 
\theta^y_i  = B_0  x_i a  \quad  \theta_{i }^x  =  B_0 Ly_i \delta_{x_i ,L}  .
\label{PeierlsPhasesClosed:eq:app:stab}
\ee
Here $x_i$ and  $y_i$  denote the coordinates of site  $i $ (defined with branch cut outside $S_0$), and $\delta_{ij}$ denotes the Kronecker delta symbol, such that $\delta_{x_i ,L} $ takes the value $1$ if $x_i =L$,  while $\delta_{x_i ,L}=0$ for all other values of $x_i $.
Recall that $a$ is the lattice constant. 
The phases $\theta^y_i $ ensure that a  trajectory  encircling a plaquette acquires a phase of  $B_0a^2$, if the trajectory does not cross the branch cut of the $x$-position operator between $x = L$ and  $x = 0$. 
The phase $\theta^x_i $, which does not appear in the Landau gauge in an open geometry, is necessary to ensure that the phase is also given by $B_0a^2\, (\mod 2\pi)$ for trajectories encircling plaquettes across the branch cut.

The goal of the following is to show that we can choose another gauge where $B_0$ only weakly perturbs the Hamiltonian within a particular finite region of the lattice, $S$, which   consists of one or more non-overlapping disk-shaped regions, $S_1,\ldots S_N$, whose combined area, $A_S $,  is much smaller than $L^2$.
We reach such a gauge through the following  transformation to the one-flux Hamiltonian with the gauge choice as prescribed in Eq.~\eqref{eq:JPeierlsModification}: $\hat c_i \to e^{-i \phi_i}\hat c_i$, where $\phi_i  = B_0 x_0^{(n)} y_i$ for sites $i$ within subregion $S_n$, and  $(x^{(n)}_0,y_0^{(n)})$ denotes the center of subregion $S_n$.
In this case, one can verify that, for sites within subregion $n$ the  Peierls phases resulting from this transformation take the following values:, 
\be
\theta_i^y = B_0(x_i-x_0^{(n)}), \quad \theta_i^x = 0.
\ee
The later holds since the branch cut of the $x$-coordinate does not intersect $S$.
Since $S_n$ has disk geometry and is centered around $(x_0^{(n)},y_0^{(n)})$, we thus find $|x_i-x_0^{(n)}|\leq \sqrt{A_S}$ for sites $i$ within subregion $S_n$.
Hence we confirm that the Peierls phases are all of order $\sqrt{A_S} a/L^2$ for bonds within $S$, and thus much smaller than $1$ in the limit $A_S \ll L^2$ specified above.

\newcommand{\res}[1]{#1}
\subsection{Response of the Hamiltonian}
\label{ResponseOfHtoFlux:sec:app:stab}

An important result  we will use extensively in the following is that, for large systems, the insertion of the uniform field $B_0$ only weakly perturbs the system, up to a gauge transformation. 
To see this, we  consider the action of  the perturbation induced by  $B_0$, $\delta H(t)\equiv \tH(t)-H(t) $ (in the particular gauge we consider),  on a  state $|\psi\rangle$ with an arbitrary number of particles, where all particles are located in the   finite region $S$ that was introduced in the previous subsection.

As a first step, we note that  $\delta H(t) |\psi\rangle =\delta H(t)P_S |\psi\rangle$, where $P_S$ projects into the subspace where all particles are located within $S$. 
Using that $\hat c_i P_S = 0$ if site $i$ is located outside $S$, we find 
\be 
\delta H(t)P_S=\sum_{j  \in S } \sum_i J_{ ij }(t)\hat c^\dagger_i  \hat c_j  (e^{i \theta_{i j } }-1) .
\ee
The Peierls phases $\{\theta_{i j }\}$ are as given in Eq.~\eqref{PeierlsPhasesClosed:eq:app:stab} above.
Below, we establish an upper bound for
the spectral norm~\cite{FN:OperatorNorm} of $\delta H(t) P_S$, $\norm{\delta H(t)P_S}$.
To do this, we use that $\norm{M}\leq \sqrt{\Tr(M^\dagger M)}$, such that 
\be 
\norm{\delta HP_S}^2 \leq\sum_{j _1, j_2\in S } \sum_{ i _1, i_2} K^*_{i_1 j_1 } K_{i _2j _2} \Tr(\hat c^\dagger_{j _1} \hat c_{i _1} \hat c^\dagger_{i _2}\hat c_{j _2}), \notag
\ee
where $K_{i j }\equiv J_{i j } (e^{i \theta_{i j } }-1)$, and we suppressed time-dependence for brevity. 
Since $\theta_{i j}=0$ for $i = j $,  terms above are only nonzero when $i _1=i _2 $ and $j _1=j _2$. 
Thus, 
\be 
\norm{\delta HP_S}^2 \leq   \sum_{j \in R} \sum_i|J_{i j }|^2 |e^{i \theta_{i j }}-1|^2.
\label{tHminusHSquared:eq:app}
\ee
 
We  now estimate the maximal scale of the right hand side above. 
We recall from the discussion in the end of Subsection~\ref{MagneticFluxImplementation:sec:app:stab} that 
the  Peierls phases $\{\theta_{i j }\}$, as given in Eq.~\eqref{PeierlsPhasesClosed:eq:app:stab}, are of order $\sqrt{A_S} a /L^{2}$ or smaller for bonds within the region $S$.
This implies that  the value of each non-vanishing term in the sum in Eq.~\eqref{tHminusHSquared:eq:app} is of order $J^2 A_S a^2/L^4$ or less, where $J$ denotes the typical scale of the (off-diagonal) tunneling coefficients $\{J_{i j }\}$. 
To estimate the number of non-vanishing terms in the sum we recall, from the assumptions made in the beginning of subsection~\ref{MagneticFluxImplementation:sec:app:stab}, that the tunneling coefficients $J_{i j }$ only couple nearest-neighbor pairs of sites in the lattice. 
Hence,  for each choice of the index $j $,  $J_{i j }$ may only be non-vanishing for four choices 
 of the index $j$. 
These considerations show that there are only of order $A_R/a^2$ non-vanishing terms in the sum above. 
Using that  each non-vanishing term has norm of order $\lesssim J^2 A_R a^2/L^4$, we find that $\norm{\delta H P_S} ^2 \lesssim A_S^2 J^2 L^{-4}$. 
Here $a\lesssim b$ indicates that $a$ is smaller than $b$, or of order $b $. 
Thus  we conclude that
\be 
\norm{\delta HP_S} \lesssim  J A_S / L^2. 
\label{HamiltonianInequality:eq:app:stab}
\ee
In the sense of the operator norm, the difference between the Hamiltonians with and without one flux quantum uniformly piercing the entire torus decays to zero with the inverse of the total system area, when acting on states confined to the region $S$, and with a judicious choice of gauge.

\subsubsection{Action on a localized state}
\label{sec:ActionOnALocalizedState}
Using the above result, we now  show that a gauge exists where $\delta H$ is small when acting on  states which are not strictly confined to the 
region $S$ of the lattice, but rather only exponentially localized.
Specifically, we consider a state $|\psi\rangle$, whose full support is exponentially confined to a region $S$ which consists of one or more disk-shaped subregions of  radius $r$,
 with the probability of finding a particle a distance $s$ from the center of the nearest subregion decaying as $e^{-s/\xi_l}$ when $s>r$. 

To conveniently quantify the extent to which particles are confined within a subregion of the lattice, for each $j=1,2\ldots $, we let $|\psi_j\rangle$ denote the component of the wavefunction $|\psi\rangle$ where the outermost particle is located in the distance interval  between $(j-1)a$ and $ja$ from the nearest subregion  of $S$.
Specifically, $|\psi_j\rangle \equiv (P_j-P_{j-1})|\psi\rangle$, where 
  $P_j$ denotes the projector onto the states where {\it all} particles are located within a distance $ja$ from the center of the nearest subregion of $S$.  
From this definition one can verify  that  $|\psi\rangle = \sum_{j=1}^\infty |\psi_j\rangle$. 
Moreover, the using that $P_j P_{k} = P_{\min(j,k)}$, it follows  that  the components are mutually orthogonal: $\langle\psi_j|\psi_k\rangle = 0$ for $j\neq k$.
%
From the definitions above, the probability for finding  finding a particle more than a distance $ja$ from the center of $R$ is given by $\langle\psi|(1-P_{j})|\psi\rangle = \sum_{j'=j+1}^\infty \langle \psi_{j'}|\psi_{j'}\rangle$. 
Since the left hand side must be of order $e^{-ja/\xi}$ for $ja >r$, and each term in the right hand side is positive, we must have
%
\be 
\langle \psi_j|\psi_j\rangle\lesssim e^{-ja /\xi_l} \quad {\rm
for} \quad j>r/a.
\label{eqa:psijnorm}
\ee


We now use the above result to obtain a bound for the state $\delta H|\psi\rangle$. 
Inserting $|\psi\rangle = \sum_{j=1}^\infty|\psi_j\rangle$, and using $P_j |\psi_j\rangle  = |\psi_j\rangle$ one can verify that 
$ 
|\psi\rangle = P_R |\psi\rangle +  \sum_{j>r/a}P_j |\psi_j\rangle,
$ where $P_S \equiv P_{r/a}$ denotes the projector into the subspace where all particles are located within the region $S$ (for convenience we assume $r$ to be an integer multiple of the lattice constant $a$). 
Using this result along with the triangle inequality and Eq.~\eqref{eqa:psijnorm}, 
we hence obtain:
\be 
\norm{\delta H|\psi\rangle} \lesssim   \norm{\delta HP_R}
 + \sum_{j>r/a} \norm{\delta HP_j} e^{-\frac{ja}{2\xi_l}}. 
\notag 
\ee
The considerations from Sec.~\ref{ResponseOfHtoFlux:sec:app:stab} show that  we may choose a gauge for $\tH$ such that $\norm{\delta H P_S}\lesssim J A_S / L^2$, and  
 $\norm{\delta HP_j }\lesssim A_{ S_j}^2 J /L^2$ for any choice of $j$, where $A_{S_j} \sim (ja)^2$ denotes the area of the region projected into by $P_j$.
Using that $\sum_{j>j_0} j^2 e^{-j/k} \sim   j_0^2 e^{-j_0/k}$ when $j_0\gg k$,  one can then verify that 
\be 
\sum_{j>r/a} \norm{\delta HP_j} e^{-\frac{ja}{2\xi_l}} \lesssim A_S J/L^2 e^{-r/2 \xi_l},
\ee
where $A_S \sim r^2$ denotes the area of the region $S$.
Thus, since $r\gg \xi_l$, we find 
\be 
\norm{\delta H|\psi\rangle} \lesssim   J A_S / L^2 .
\label{eq:LocalizedStateResult}
\ee
\subsection{Response of the Floquet operator}
\label{ResponseOfUToFlux:sec:app:stab}
We now show that, for any region $S$ in the lattice that consists of one or more disk-shaped subregions,  it is possible to find a gauge, the  Floquet operators of the one- and zero-flux systems, $\tU(T)$ and $U(T)$,  have nearly identical actions  states $|\psi\rangle$  localized within $S$: $\tU (T)|\psi\rangle \approx U(T)|\psi\rangle$. 
 Here the state is said to be localized within $S$ if  the probability of finding a particle a distance $s$ from the center of the nearest subregion os $S$ decays as $e^{-s/\xi_l}$ for $s>r$, where $r$ denotes the radius of $S$. 

First, we note that $\lVert{(U-\tU)|\psi\rangle\rVert} = \lVert{(\tU^\dagger U - 1)|\psi\rangle}\rVert$. 
This follows from the unitarity of $\tU$,  since for any state $|\Psi\rangle$, $\lVert{|\Psi\rangle}\rVert=\lVert{\tU^\dagger|\Psi\rangle}\rVert$. 
Using  that 
$ 
\tU^\dagger U - 1 = \int_0^T \!\! {\rm d}t\, \partial _t [\tU^\dagger(t) U(t)]
$,
along with $\delta H(t) \equiv \tilde H(t) - H(t)$, 
we find 
\be 
(U-\tU)|\psi\rangle =-i\! \int_0^T \! \!\! {\rm d}t\, \tU^\dagger(t) \delta H(t) U(t)|\psi\rangle.
\label{eq:W(t)Result}
\ee
Using that $\lVert{|\Psi\rangle}\rVert=\lVert{\tU^\dagger|\Psi\rangle}\rVert$  along with the triangle inequality, we thus find 
\be 
\lVert(U-\tU)|\psi\rangle\rVert \leq  \int_0^T\!\!\! {\rm d}t\,   \norm{  \delta H(t) U(t) |\psi\rangle}.
\label{eq:EqB13}
\ee

We now  use that $U(t)$ is local at  all times $0\leq t \leq T$, due to the finite Lieb-Robinson velocity $v$ of the system. 
The locality implies that, for  the state $U(t)|\psi\rangle$, the probability of finding a particle a distance $s$ from the center of $S$   decays exponentially when  $s\gtrsim r$. 
Using the result in Eq.~\eqref{eq:LocalizedStateResult} from the previous subsection, we thus find 
\be
\norm{\delta H(t)U(t)|\psi\rangle} \lesssim J A_S/L^2. 
\label{eq:HUPsi_inequality}
\ee
Using this in the inequality in~Eq.~\eqref{eq:EqB13}, we conclude 
\be 
\norm{(U^\dagger \tU -1) |\psi\rangle} \lesssim  J TA_S /L^2.
\label{eq:FloquetOperatorResult}
\ee
Thus, $\norm{(\tU -U) |\psi\rangle} \lesssim  J TA_S /L^2$.

The  result in Eq.~\eqref{eq:FloquetOperatorResult} shows that, with a judicious choice of gauge, the Floquet operators of the one- and zero flux systems  give nearly identical results when acting on a localized state.
In this sense, the insertion of a uniform magnetic field $B_0$ only weakly modifies the Floquet operator for large systems.

\subsection{Response of Floquet eigenstates and quasienergy spectrum}
\label{FloquetEigenstateCorrespondence:sec:app:stab}
\newcommand{\as}{almost surely }
We now show that, {in the subspace with $k$ or fewer particles}, the quasienergy spectrum and Floquet eigenstates of $k$-particle localized systems are robust to perturbations, and only weakly affected by the insertion of the uniform magnetic field $B_0$.

In this subsection, it is useful to  use   notation that  relates the quasienergies and Floquet eigenstates to the LIOM decomposition in Eq.~\eqref{eq:FloquetOperatorForm} (which is valid  in the subspace of up to $k$ particles, which we consider): 
in the following we thus let $|\Psia\rangle \equiv \hat f^\dagger_{\alpha _1} \ldots \hat f^\dagger_{\alpha _\ell}|0\rangle$  denote the Floquet eigenstate of the system for which only LIOMs $\al$ take value $1$  (see Sec.~\ref{sec:liom_structure} for definition of  $\hat f^\dagger_\alpha $), and  let $E_{\al}$ denote the corresponding quasienergy. 



Using this cutoff length, we show below that for each finite $\ell \leq k$, where $k$ denotes the system's degree of localization (which is infinite for MBL systems), the $\ell$-particle Floquet eigenstates $\{|\tilde \Psi_\al\rangle\}$ of $\tU$ can be labeled such that, for {\it each} choice of LIOMs   (identified by the LIOM indices $\al$), 
\be 
|\tilde \Psi_\al\rangle = |\Psia\rangle + \mathcal O\left (L^{-1/2}\right),
\label{eqa:fes_cor_1}
\ee
and 
\be 
\tilde E _\al = E _\al + \mathcal O \left( L^{-2}\right).
\label{eqa:qe_cor_1}
\ee
Eq.~\eqref{eqa:fes_cor_1}  thus shows that, in the thermodynamic limit, {\it each} eigenstate of $\tU$ is identical to an eigenstate of $U$, up to gauge transformation and a vanishingly small correction, while Eq.~\eqref{eqa:qe_cor_1} shows that their associated quasienergies similarly are identical up to  a vanishing correction.
This establishes the one-to-one correspondence of the quasienergy levels of the zero- and one-flux systems that we summarized below Eq.~\eqref{eqa:qel_relation}.

Due to the possibility that the 
 field $B_0$ induces a resonance between two Floquet eigenstates of $U$, disorder realizations do exist where one (or more) of the eigenstates of $\tU$ is a significantly hybridized combination of two eigenstates of $U$. 
In this case,  Eq.~\eqref{eqa:fes_cor_1} will hold for most but not all Floquet eigenstates of the system.
However, as we show here,  the set of disorder realization where such a resonance-induced breakdown of Eq.~\eqref{eqa:fes_cor_1} occurs  has measure zero in the thermodynamic limit. 
In this way, Eqs.~\eqref{eqa:fes_cor_1}~and~\eqref{eqa:qe_cor_1} hold for {\it almost all} disorder realizations, in the thermodynamic limit.  

To establish Eqs.~\eqref{eqa:fes_cor_1}~and~\eqref{eqa:qe_cor_1}, we first consider the case $\ell=1$ (i.e., we establish the relationships for each single-particle Floquet eigenstate).
Subsequently, 
in a stepwise fashion, we generalize this result to states with $\ell$ particles, for each $\ell = 2,\ldots k$. 

\subsubsection{Single-particle eigenstates }
\label{sec:SingleParticleStates}
Here we establish the relationships in Eqs.~\eqref{eqa:fes_cor_1}~and~\eqref{eqa:qe_cor_1}  for the single-particle case.
We assume that $k$-particle localization is robust to perturbations, and thus $\tU$ also describes a $k$-particle localized system (we assume $k\geq 1$).
Thus,  in particular,  each single-particle eigenstate $|\tilde\Psi\rangle$ of $\tU$  has its  full support within a finite disk-shaped region $S$ of linear dimension $d$,
with the probability of finding the particle a distance $s$ outside $S$ decaying as  $e^{-s/\xi_l}$. 

Due to its finite region of support, each single-particle eigenstate of $\tU$, $|\tilde \Psi\rangle$,  may only overlap  significantly with Floquet  eigenstates whose corresponding LIOM centers are located within a distance $\sim \xi_l$  from $S$.
To exploit this fact, we introduce  a system-size dependent length scale $d\gg \xi_l$, which acts as an effective length cutoff for the region of support of a LIOM. 
The length $d$ must be much smaller than $L$, but can otherwise  be taken to be arbitrarily large, as long as $d/ L$ vanishes in the thermodynamic limit.  
From the considerations above it follows  that $|\tilde \Psi\rangle$  only overlaps with the finite number Floquet eigenstates, $|\Psi_{\alpha _1}\rangle\ldots |\Psi_{\alpha _{N_1}}\rangle$, whose LIOM centers are located within a distance $d$ from $S$, (up to a correction exponentially small in $d/\xi_l$:
\be 
 \sum_{n=1}^{N_1}| \langle \Psi_{\alpha _n}|\tilde \Psi\rangle|^2  = 1 + \mathcal O (e^{-d/\xi_l}). 
 \label{eq:PsiPsiTildeSum}
\ee
For the purposes of the following, it is convenient to order the indices $n$  according to the value of the overlap, such that $| \langle \Psi_{\alpha _1}|\tilde \Psi\rangle|^2 \geq | \langle \Psi_{\alpha _2}|\tilde \Psi\rangle|^2 \geq \ldots \geq | \langle \Psi_{{\alpha _{N_1}}}|\tilde \Psi\rangle|^2 $. 
Note that the sequence of LIOM indices $\alpha _1 \ldots \alpha _{N_1}$  depends on the choice of $|\tilde \Psi\rangle$; this dependence is taken to be implicit below,  for the sake of brevity.

We now show that $|\tilde \Psi\rangle$  only overlaps significantly with {\it one} of the eigenstates $|\Psi_{\alpha _1}\rangle \ldots |\Psi_{\alpha _{N_1}}\rangle$, while the total weight from all other eigenstates gives a negligible contribution. 
To show this, note that $|\Psi_{\alpha _n}\rangle$ and $|\tilde \Psi\rangle$ are eigenstates of $U$ and $\tU$, respectively, and hence
\be 
\langle \Psi_{\alpha _n} |\tilde \Psi\rangle = \frac{\langle  \Psi_{\alpha _n} |  U^\dagger \tU-  1 |\tilde \Psi\rangle }{e^{-i(\tilde E- E_{\alpha _n} )T}-1},
\label{eq:QuasienergyPsiRelation}
\ee
where $\tilde E$  is the quasienergy associated with $|\tPsi\rangle$. 
Since  $|\tilde \Psi\rangle$ is exponentially well localized within 
 $S$, Eq.~\eqref{eq:FloquetOperatorResult}  implies that   $|\langle \Psi_{\alpha _n} | U^\dagger \tU-  1 |\tilde \Psi\rangle| \lesssim JT A_S / L^2$. 
Moreover, $|e^{-i(\tilde E - E_{\alpha _n})T}-1|\leq |\tilde E- E_n|T$, where the  norm  $|\cdot|$ is defined modulo $2\pi/T$, i.e. $|E|\equiv \min_z |E +2\pi z/T|$.
Combining these two inequalities with Eq.~\eqref{eq:QuasienergyPsiRelation}, we find 
\be 
|\langle \Psi_{\alpha _n} |\tilde \Psi\rangle|\lesssim \frac{ J A_S/L^2T}{|\tilde E - E_{  \alpha _n}| }.
\label{eq:FractionResult}
\ee

We now consider two implications of the above inequality. 
Firstly,  Eq.~\eqref{eq:PsiPsiTildeSum} implies $|\langle \Psi_{\alpha_1} |\tilde \Psi\rangle|^2\gtrsim 1/N_1-\mathcal O(e^{-d/\xi_l})$ (c.f. the labelling of the states $\{|\Psi_{\alpha _n}\rangle\}$). Thus, 
\be  
|\tilde E-E_{\alpha _1}| \lesssim {\sqrt{N_1} J A_S}/{L^2 T}.
\label{eq:Ealpha1Result}
\ee
Secondly, we note that, for a random choice of $|\tilde \Psi\rangle$, the typical spacing between the $N_1$ quasienergy levels $\{E_{n}\}$ is of order $\Delta E \sim \mathcal W/N_1$, where $\mathcal W$ denotes the  width of the single-particle quasienergy spectrum (when the quasienergy spectrum has no gaps, $\mathcal W = 2\pi/T$).
In this case, only one of the quasienergies $\{E_{\alpha _n}\}$ (namely $E_{\alpha _1}$)  is  close enough to $\tilde E$ for Eq.~\eqref{eq:FractionResult} to allow a significant value of  $\langle \Psi_n|\tPsi\rangle$. 
Thus, $|\tPsi\rangle \approx |\Psi_1\rangle$ for a typical choice of  $|\tilde \Psi\rangle$.

We now prove that $|\tPsi\rangle\approx |\Psi_1\rangle$ for {\it any} choice of $|\tPsi\rangle$ in the system (except for a measure-zero set of disorder realizations in the thermodynamic limit). 
To establish this result, we first note 
\be 
|E_{n}-\tilde E|\geq |E_{\alpha _n}-E_{\alpha _1}|-|\tilde E-E_{\alpha _1}| .
\label{eq:TriangleInequality}
\ee  
We now establish a lower bound for $ |E_{\alpha _n}- E_{\alpha _1}|$, using the fact the quasienergy levels of nearby states $E_{\alpha _1} $ and $E_{\alpha _n}$   repel each other, and that $|\tilde E-E_{\alpha _1}|$ satisfies the bound of Eq.~\eqref{eq:Ealpha1Result}.
Specifically,   note that the   Floquet eigenstates $|\Psi_1\rangle$ and $|\Psi_{n}\rangle$ have their support within a distance $\lesssim d$ from each other. 
The quasienergies $E_{\alpha _1}$ and $E_{\alpha _n}$ are hence subject to local level repulsion when the quasienergy difference $\delta E  \equiv |E_{n}-E_{1}|$ is  much smaller than the scale of matrix elements between them with respect to the kinetic part of the Hamiltonian (i.e. $\delta E\ll J e^{-d/\xi_l}$). 
In the limit where $\delta E \ll J e^{-d/\xi_l}$, 
 the   probability distribution $p(\delta E)$ for $\delta E$ should thus resemble the Wigner-Dyson distribution for the Circular unitary ensemble (CUE)~\cite{RandomMatrices}: 
\be
p(\delta E) = \frac{T^3}{\pi }\delta E^2 + \mathcal O (\delta E^4).
\ee

Using the above result, we  now compute the expected number of  pairs of nearby single-particle eigenstates $|\Psi_{\alpha _i}\rangle$ and $|\Psi_{\alpha _j}\rangle$  in the entire system, for which
 $|E_{\alpha_i}-E_{\alpha_j}|$ is smaller than some given (small) value $\delta E_0$. 
 Here ``nearby'' refers to the eigenstates $|\Psi_{\alpha _i}\rangle$ and $|\Psi_{\alpha _j}\rangle$  having their centers located within a distance $\sim d$ from each other, such that they may potentially overlap with the same eigenstate of $\tU$. 
Noting that there are $\mathcal O (L^2 N_1/2a^2)$ distinct   pairs of nearby eigenstates (where $a$ denotes the lattice constant), we have 
\be
N(\delta E_0)= \frac{L^2 d^2}{2a^4} \int_0^{\delta E_0} \!\!\! d\delta E\, p(\delta E) .
\ee
where we used $N_1\sim (d/a)^2$.
Thus, in the limit where $\delta E_0\ll J e^{-d/\xi_l}$, 
\be
N (\delta E_0) =  \frac{L^2 d^2(\delta E_0 T)^3}{6 \pi a^4}.
\label{eqa:ndeltae0}
\ee

We recall we may take $d$ arbitrarily large  as long as $d/L \to 0$ in the thermodynamic limit. 
In the following, it is convenient to let $d$ scale with system-size as $d \sim \frac{1}{2}\xi_\ell \log(L/a)$ such that $\mathcal O(e^{-d/\xi_l}) \sim \mathcal O(L^{-1/2})$. (Note that this choice is not unique; other   scaling behaviors of $d$ can be  used in the discussion below).
Since with this choice of $d$, $Je^{-d/\xi} \gg a/LT$ in the thermodynamic limit [such that Eq.~\eqref{eqa:ndeltae0} applies to  $\delta E_0 = a/LT$], we conclude that 
\be
\lim_{L\to \infty} N( a/LT) = 0.
\ee
We conclude  that, in the thermodynamic limit, there are zero pairs of Floquet eigenstates $|\Psi_{\alpha _i}\rangle$ and $|\Psi_{\alpha _j}\rangle$ with LIOM centers within a distance $d \sim \frac{1}{2}\xi_\ell \log(L/a)$ from each other whose quasienergies differ by less than $\frac{a}{LT}$ (except for in a measure zero set of disorder realizations).
We conclude, in the thermodynamic limit, and for {\it any} choice of $|\tilde \Psi\rangle$, 
\be 
| E_{\alpha _1}-E_{\alpha _n}| > \frac{a }{LT}
\ee
for all but a measure zero set of disorder realizations.

Using Eq.~\eqref{eq:TriangleInequality} along with the fact   that $|\tilde E - E_{\alpha _1}|$ is subleading in $L$ compared to the above bound for $|E_{\alpha _n}-E_{\alpha _1}|$,  we find, for $n\geq 2$, $ 
|\tilde E-E_{\alpha _n}|> \frac{a}{ L T}.
$
Thus, for all but a measure zero set of disorder realizations, it holds that, for {\it each} choice of $|\tilde \Psi\rangle$, 
\be 
|\langle \Psi_{\alpha_n} |\tilde \Psi\rangle|  < \frac{ A_S JT  }{ a L } \quad{\rm for} \quad n\geq 2.
\ee
Using this result in Eq.~\eqref{eq:PsiPsiTildeSum}, we find
\be 
1-|\langle \Psi_{\alpha _1}|\tilde \Psi\rangle|^2<  \frac{N_1 A_S^2 J^2 T^2   }{ a^2 L^2}+\mathcal O (e^{-d/\xi_l}).
\ee
Recall that we take $d\sim \frac{1}{2}\xi_l \log(L/a)$. Hence the first term above is subleading in the thermodynamic limit, and we obtain
\be 
|\langle \Psi_{\alpha _1}|\tilde \Psi\rangle|^2  =1 +\mathcal O (L^{-1/2}) . 
\ee
(See Footnote~\onlinecite{fn:correction_def}).
This concludes the proof of Eq.~\eqref{eqa:fes_cor_1} for the single-particle case, when we assign the label $\alpha _1$ to $|\tPsi\rangle$.

To establish Eq.~\eqref{eqa:qe_cor_1} for the single-particle case, we note from Eq.~\eqref{eq:FractionResult} [with the labelling for $|\tilde \Psi\rangle$ introduced above] that, for each choice of $\alpha $, 
\be 
|\tilde E_{\alpha}  - E_{\alpha}|
\lesssim \frac{ J A_S/L^2}{|\langle \tilde\Psi_{\alpha} |\Psi_\alpha \rangle| }.
\ee
Since $|\langle \tPsi_\alpha |\Psi_\alpha \rangle|\approx 1$, and $A_S \sim d^2$, we conclude $|\tilde E_\alpha  - E_{\alpha}| \lesssim J d^2/L^2 \sim \mathcal O (L^{-2})$ (see Footnote~\cite{fn:correction_def}).
This is what we wanted to show.

\subsubsection{Two-particle eigenstates}
Having established  Eq.~\eqref{eqa:fes_cor_1}  for single-particle Floquet eigenstates, we now show that it also holds for all two-particle eigenstates (provided the system is $k$-particle localized for some $k\geq 2$).
In order to do this, we consider a two-particle   Floquet eigenstate $|\tilde\psi\rangle$ of the  one-flux system, with quasienergy $\tilde E$.
Since the one-flux system is $k$-particle localized (for $k\geq 2$), the two-particle  eigenstates of $\tU$ possess a LIOM structure.
In the Floquet eigenstate $|\tilde\psi\rangle$, two of the LIOMs of $\tU$, $\tilde n_1$ and $\tilde n_2$, are thus ``excited'' (i.e. $\tilde n_{\alpha }|\tilde\psi\rangle= |\tilde\psi\rangle$ for $\alpha =1,2$, while $\tilde n_{\alpha }|\tilde\psi\rangle= 0$ for  $\alpha \neq 1,2$). 
In the following, we  divide our argumentation into two cases, depending on whether or not the  LIOMs $\tilde n_1$ and $\tilde n_2$  are located within a distance $d$ from each other, where $d$ denotes the arbitrary  length scale cutoff  for each LIOM's region of support introduced in Sec.~\ref{sec:SingleParticleStates}.
\paragraph*{Nearby LIOMs ---}
When the centers of the two ``excited'' LIOMs $\tilde n_1$ and $\tilde n_2$ in the state $|\tilde\psi\rangle$ are separated by a  distance less than $d$,  a two-particle Floquet eigenstate $| \Psi_{\alpha \beta }\rangle$ of the zero-flux system may only  significantly overlap with $|\tilde\Psi\rangle$ if the corresponding excited (nonperturbed) LIOMs $\nhat_\alpha $ and $\nhat _\beta $ are  located within a distance $d$ from the centers of $\tilde n_1$ and $\tilde n_2$.
As a result, there are only of order $N_2 \sim \binom{2 d^2/a^2}{2}$ 
 choices of distinct LIOMs $\alpha ,\beta $  for which $| \Psi_{\alpha \beta }\rangle$ can significantly overlap with $|\tilde\Psi\rangle$.

Using the  same arguments as for the single particle case (Sec.~\ref{sec:SingleParticleStates}) one can show that, for all but a measure-zero set of disorder realizations in the thermodynamic limit,  there exists a unique two-particle eigenstate $| \Psi_{\alpha \beta }\rangle$ of $U$ for each two-particle eigenstate $|\tilde \Psi\rangle$ of $\tU$ such that (up to a gauge transformation) 
\be 
|\tilde \Psi\rangle = |\Psi_{\alpha \beta } \rangle + \mathcal O \left (  L^{-1/2}\right),
\ee
and 
\be 
\tilde E = E_{\alpha \beta } + \mathcal O \left(L^{-2}\right).
\ee

\paragraph*{Separated LIOMs ---} 
Next, we consider  the case where the two excited LIOMs $\tilde n_1$ and $\tilde n_2$ 
 are separated by a  distance $\Delta r$ larger than $d$. 
In this case, the LIOM structure of the Floquet operator $\tU$ [Eq.~\eqref{eq:FloquetOperatorForm} in the main text] implies that, up to an exponentially small correction in the distance $\Delta r /\xi_l$, 
 $|\tilde \Psi\rangle $ may be written as a direct product of two single-particle eigenstates 
$|\tilde \Psi_{\alpha }\rangle $ and $|\tilde \Psi_{\beta }\rangle$. 
Here $\alpha $ and $\beta $ refer to the labeling of the single-particle eigenstates of $\tU$ that was established in the previous subsection. 
Letting  $S_\alpha $  and $S_\beta $ denote  the two non-overlapping regions of linear dimension $d$ where the states $|\tilde \Psi_\alpha \rangle$ and $|\tilde \Psi_\beta \rangle$ 
respectively have their support (up to a correction exponentially small in $d/\xi_l$), we have~\cite{FN:TensorProuctDefinition}:
\be 
|\tilde \Psi\rangle = |\tilde \Psi_{\alpha }\rangle_{S_{\alpha }} \otimes |\tilde \Psi_{\beta }\rangle_{S_{\beta }} \otimes | 0\rangle
 + \mathcal O(e^{-d/\xi_l}).
 \label{eq:app:stab:Two-particleFloquetStatesTensorProduct}
\ee 
where we used $\Delta r > d$.
Here $| \Psi\rangle_{S}$  denotes the restriction of the state $|\Psi\rangle$ to the Fock space of the  region $S$ (defined from the projection of  $|\Psi\rangle$ into the subspace with no particles outside region $S$). 
The state  $ | 0\rangle$  refers to the vacuum in the complementary region to $S_{\alpha }$ and $S_{\beta }$. 
Since the two particles in the state $|\tilde \Psi\rangle$  are separated by a distance much larger than $d$, the regions $S_{\alpha }$ and $S_{\beta }$  do  not overlap. 

We recall that Eq.~\eqref{eqa:fes_cor_1} was already proven to hold for the single-particle case. 
Thus  $|\tilde \Psi_{\alpha }\rangle$ (the eigenstate in the presence of one flux quantum piercing the system) is approximately identical to a single-particle eigenstate $|\Psi_\alpha \rangle$ of the zero-flux system's Floquet operator $U$ (for all but a measure zero set of disorder realizations). 
Specifically, up to a gauge transformation, $|\tilde \Psi_\alpha \rangle = |\Psi_\alpha \rangle + \mathcal O (L^{-2})$.
The eigenstate $|\Psi_\alpha \rangle$ moreover has its full support in the same  region $S_\alpha $ as $|\tilde \Psi_\alpha \rangle$, up to a correction exponentially small in $d/\xi_l$. 
Letting $V_\alpha $ be the unitary operator that generates the transformation to the gauge in which Eq.~\eqref{eqa:fes_cor_1} holds for $|\tilde \Psi_\alpha \rangle$, we have 
\be 
|\tilde \Psi_\alpha \rangle_{S_\alpha } =V_\alpha  |\Psi_\alpha \rangle_{S_\alpha }+\mathcal  O  (L^{-1/2}) ,
\label{eq:app:stab:SingleParticleRestrictedStatesRelation}
\ee
where we used that we may take $d\sim \frac{1}{2}\xi_l \log(L/a)$, such that the correction $\mathcal O (e^{-d/\xi})$ scales with system size as $L^{-1/2}$ in the thermodynamic limit. 
Using the relation~\eqref{eq:app:stab:SingleParticleRestrictedStatesRelation} for the states $|\tilde\Psi_\alpha \rangle_{S_\alpha }$ and $|\tilde \Psi_\beta \rangle_{S_\beta }$ in Eq.~\eqref{eq:app:stab:Two-particleFloquetStatesTensorProduct}, we hence obtain
\be 
|\tilde \Psi\rangle = V_{\alpha }V_{\beta } | \Psi_{\alpha }\rangle_{S_{\alpha }} \otimes | \Psi_{\beta }\rangle_{S_\beta } \otimes |0\rangle
+\mathcal  O  (L^{-1/2}) .
\ee

Due to the LIOM structure  of the Floquet operator $U$ (Eq.~\eqref{eq:FloquetOperatorForm} in the main text),  $ | \Psi_{\alpha }\rangle_{S_{\alpha }} \otimes | \Psi_{\beta }\rangle_{S_\beta } \otimes |0\rangle$ is identical to the Floquet eigenstate $|\Psi_{\alpha \beta }\rangle$ of the zero-flux system, up to a correction of order $e^{-d /\xi_l}$. 
Since the product of the two gauge transformations $V_\alpha $ and $V_\beta $ is itself a gauge transformation, we  thus conclude that, up to a gauge transformation:
\be 
|\tilde \Psi\rangle =  |\Psi_{\alpha \beta }\rangle +\mathcal  O  (L^{-1/2}) . 
\label{eq:app:stab:TwoParticleSeparatedFloquetEigenstateCorrespondence}
\ee 

The two cases we considered above show that, in the thermodynamic limit, and for all but a measure zero set of disorder realizations, {\it each} two-particle eigenstate  $|\tilde \Psi\rangle$  of $\tU$ is identical to a unique eigenstate of $U$, up to a gauge transformation, and a correction of order $\mathcal  O  (L^{-1/2})  $. 
We may thus label the two-particle eigenstates of $\tU$ such that 
Eqs.~\eqref{eqa:fes_cor_1}~and~\eqref{eqa:qe_cor_1} hold with $\ell=2$, and for each choice of the LIOM indices $\alpha_1$ and $\alpha_2$.
\subsubsection{$\ell$-particle-eigenstates}
We finally consider the general case of an $\ell$-particle eigenstate $|\tilde \Psi\rangle$ of $\tU$, where $\ell $ is smaller than or equal to the system's degree of localization, $k$.
For this situation, we  can apply the same structure of arguments as for the two-particle case:  
due to the LIOM structure of the one-flux Floquet operator $\tU$, each $\ell$-particle state is constructed by ``exciting'' $\ell$ LIOMs  $\tilde n_1 \ldots \tilde n_\ell$. 
We split our line of arguments into two cases, depending on whether or not the LIOMs $\tilde n_1 \ldots \tilde n_\ell$ can be divided into clusters separated from each other by distances greater than $d $. 
 
In the case where  the excited LIOMs {\it can} be divided into clusters in the way above, $|\tilde\psi\rangle$  can be written as  a direct product of  eigenstates of $\tU$ with fewer than $k$ particles, up to a correction of order $e^{-d/\xi_l}$. 
 Following the same line of arguments as for the analogous two-particle case,  the relationships~\eqref{eqa:fes_cor_1}~and~\eqref{eqa:qe_cor_1} can then be demonstrated to hold for this class of eigenstates using the fact that Eq.~\eqref{eqa:fes_cor_1}~and~\eqref{eqa:qe_cor_1} hold for eigenstates with fewer than $\ell$ particles.
 
In the case where all  LIOMs are located in the same cluster,  we  note that $|\tilde\psi\rangle$ only significantly overlaps  with  eigenstates $\{|\Psia\rangle\}$ where the centers of all the LIOMs $\nhat_{\alpha_1}\ldots \nhat_{\alpha_\ell}$ are located in the region $S$, consisting of all sites with a distance $d$ from any of the excited LIOM's $\tilde n_1\ldots \tilde n_\ell$. 
There only exist a finite number of eigenstates $N_\ell$ with this property.
	Specifically, $N_\ell \lesssim \binom{\ell d^2/a^2}{\ell}$ counts the number of distinct configurations of $k$ LIOMs $\nhat_{\alpha_1}\ldots \nhat_{\alpha_\ell}$ whose centers are located within $S$.
Crucially, $N_\ell$    only depends on the number of particles, $\ell$, and $d$, and is independent of system size.
 
Using the same arguments as for the single-particle case, we then find that, for all but a measure zero set of disorder realizations in the thermodynamic limit, there exists a unique eigenstate $|\Psia\rangle$ of $U$  such that (up to a gauge transformation), 
\begin{eqnarray}
|\tilde \Psi\rangle &=& |\Psi_{\al} \rangle +\mathcal O(L^{-1/2}),
\label{eq:PsiAResult}
\end{eqnarray}
where, as we  described in the beginning of this Appendix, $\mathcal O(L^{-p})$ denotes term scaling with system size as $L^{-p}$ in the thermodynamic limit (see Footnote~\cite{fn:correction_def}).
In addition, when the LIOMs are located within a distance $d$ from the same point, 
\be 
\tilde E =   E_{\al} + \mathcal O \left( L^{-2}\right).
\label{eqa:qerel}
\ee  
Thus, 
Eqs.~\eqref{eqa:fes_cor_1}~and~\eqref{eqa:qe_cor_1} hold for the $\ell$-particle case in the thermodynamic limit, for any $\ell = 1, \ldots k$.

\subsection{Relationship between magnetization density and quasienergy}
\label{MagQERelationship}

Having established the auxiliary results in Secs.~\ref{MagneticFluxImplementation:sec:app:stab}-\ref{FloquetEigenstateCorrespondence:sec:app:stab}, we are now ready to prove Eq.~\eqref{eqa:qel_relation_aide}, which is the first main goal of this appendix.
To recapitulate,  we seek to show that, for each $\ell$-particle Floquet eigenstate, $|\psi_n\rangle$ with quasienergy $\varepsilon _n$, the associated quasienergy for the one-flux system, $\tilde \varepsilon _n$ (see Sec.~\ref{FloquetEigenstateCorrespondence:sec:app:stab} for details), satisfies 
\be 
\tilde \varepsilon _n   = \varepsilon _n   +  B_0\langle \psi_n|\bar M|\psi_n \rangle+\mathcal O ( L^{-5/2}),
\label{eqa:eq00}
\ee
where $\bar M$ denotes the time-averaged magnetization operator (see Sec.~\ref{ValuesOfTheInvariants:sec} of the main text), and, as in Sec.~\ref{FloquetEigenstateCorrespondence:sec:app:stab} above, $\mathcal O(L^{-p})$  denotes a correction of order $\lambda L^{-p}$ or less, where $\lambda$ is some 
system-size independent energy scale that does not play a role for our discussion. 

In this step of the derivation  it is useful to define a region of support, $S_n$, for each Floquet eigenstate $|\psi_n\rangle$. 
{Specifically, for each Floquet eigenstate, $|\psi_n\rangle$, and for some  length scale $d \ll L$, we let $S_n$ denote the smallest region of the lattice 
that ensures the centers of all nonzero LIOMs in the state $|\psi_n\rangle$, $\al$, are located within a distance $d$ from the boundary of   $S_n$. }
The region of support $S_n$   may consist of one or   several  disconnected disk-shaped subregions of linear dimension  $~d$, and has area $A_{S_n} \leq \pi \ell d^2$. 
As in Sec.~\ref{FloquetEigenstateCorrespondence:sec:app:stab}, when taking the thermodynamic limit   $L\to \infty$ in the following, we let $d$ increase logarithmically with system size as $d\sim  \frac{1}{2} \xi_l \log( L/a) $.

To establish Eq.~\eqref{eqa:eq00}, for a given Floquet eigenstate $|\psi_n\rangle$, we let $\tU$ be the one-flux Floquet operator in a gauge where Eq.~\eqref{eq:FloquetOperatorResult} holds within   $S_n$, and let $|\tilde \psi_n\rangle$ denote the eigenstate of $\tU$ corresponding to $|\psi_n\rangle$ through Eq.~\eqref{eqa:fes_cor_1}. 
Noting that
 $|\psi_n \rangle$ and $|\tilde\psi_n \rangle$ are eigenstates of $U$ and $\tU$, respectively, we have 
\be 
\langle \psi_n |U^\dagger \tU|\tilde \psi_n \rangle = e^{-i(\tilde \varepsilon _n -  \varepsilon _n )T}\langle \psi_n |\tilde\psi_n \rangle.
\label{eq:OverlapQuasiEnergy}
\ee
At the same time, the left-hand side above  can be written [see Eq.~\eqref{eq:W(t)Result}],
\begin{align}
&\langle \psi_n |U^\dagger \tU|\tilde\psi_n \rangle =  \langle \psi_n |\tilde\psi_n\rangle \notag \\&-i \int_0^T dt \langle \psi_n |U^\dagger(t)\delta H(t)\tU(t)|\tilde\psi_n \rangle.
\label{eq:ExpansionOfUtilde}
\end{align}
We now seek to rewrite the second term above to a form which only relies  quantities of the (unperturbed) zero-flux system.
Using that  $|\tilde \psi_n \rangle = |\psi_n \rangle +\mathcal O(L^{-1/2})$ [Eq.~\eqref{eqa:fes_cor_1}], and  that $U|\psi\rangle = \tU |\psi\rangle + \mathcal O(L^{-2})$ for normalized states $|\psi\rangle$ that are exponentially localized within $S_n$  (such as $|\psi_n\rangle$), we find 
\be 
\tU(t)|\tilde\psi_n \rangle = U(t)|\psi_n\rangle + \mathcal  O(L^{-1/2}).
\label{eq:TildeToNoTilde}
\ee
We  recall from Eq.~\eqref{eq:HUPsi_inequality} that $\norm{|\delta H(t)U(t)|\psi_n\rangle}\sim \mathcal O (L^{-2})$, such that, for any state $|\psi\rangle$, $|\langle \psi_n |U^\dagger(t)\delta H(t)|\psi\rangle| \lesssim \mathcal O(L^{-2}) \norm{|\psi\rangle}$.
Combining this with Eqs.~\eqref{eq:ExpansionOfUtilde}~and~\eqref{eq:TildeToNoTilde}, we find
\begin{align}
&e^{-i(\tilde \varepsilon _n-  \varepsilon _n)T}\langle \psi_n|\tilde\psi_n\rangle
 =  \langle \psi_n|\tilde\psi_n\rangle \notag \\&-i \int_0^T \!\!\!\!dt \langle \psi_n|U^\dagger(t)\delta H(t)U(t)|\psi_n\rangle +\mathcal O(L^{-5/2}).
\end{align}
We finally note that $\langle \psi_n|\tilde\psi_n\rangle = 1+\mathcal O(L^{-2})$. 
Dividing through with a factor of $\langle \psi_n|\tilde\psi_n\rangle $, and again using that $\norm{\langle \psi_n|U^\dagger(t)\delta H(t)}~\sim \mathcal O(L^{-2})$, we hence obtain
\begin{align}
e^{-i(\tilde \varepsilon _n-  \varepsilon _n)T} &= 1-i \int_0^T \!\!\!\!dt \langle \psi_n|U^\dagger(t)\delta H(t)U(t)|\psi_n\rangle \notag \\
&+\mathcal O(L^{-5/2}).
\label{eq:Step45}
\end{align}
Expanding the left-hand side to first order in $\tilde \varepsilon _n- \varepsilon _n$, and using $\tilde \varepsilon _n -\varepsilon _n \sim \mathcal O(L^{-2})$ [see Eq.~\eqref{eqa:qerel}], we obtain
\begin{align}
\tilde \varepsilon _n-  \varepsilon _n = &\frac{1}{T}\int_0^T \!\!\!\! dt \langle \psi_n|U^\dagger(t)\delta H(t)U(t)|\psi_n\rangle +\mathcal O(L^{-5/2}).
\label{eq:Step4512}
\end{align}

Having expressed $\tilde \varepsilon _n - \varepsilon _n$ purely in terms of quantities of the zero-flux system, we now relate the first term on the right-hand side above to the time-averaged magnetization in the state $|\psi_n\rangle$. 
To this end, we  use  the explicit form of $H(t)$ we assumed in Eq.~\eqref{eq:app:Hamiltonian:stab} (similar arguments apply to more general Hamiltonians), finding 
\be 
\delta H(t) =  i \sum_{ij} \theta_{ij} J_{ij}(t)\hat c^\dagger_i \hat c_j +  \delta H^{(2)}(t),
\ee
where $\delta H^{(2)}(t) = \sum_{ij} [e^{i\theta_{ij}}-1-i\theta_{ij}]J_{ij}(t)\hat c^\dagger_i \hat c_j$ and $\{\theta_{ij}\}$ denote the Peierls phases induced by the uniform magnetic field $B_0$. 
We identify $-i  [J_{ij}(t) \hat c^\dagger_i \hat c_j - J_{ji}(t) \hat c^\dagger_j \hat c_i]$ as the bond current operator on the bond from site $j$ to site $i$, $\hat I_b(t)$, and $\theta_{ij}$ as the associated Peierls phase  (see also Footnote~\onlinecite{fn:bond_orientation}).
Hence,    $ i \sum_{ij} \theta_{ij} J_{ij}(t)\hat c^\dagger_i \hat c_j  = -\sum_b \theta_b I_b(t)$.

To establish a bound for the term in Eq.~\eqref{eq:Step4512} originating from $\delta H^{(2)}(t)$, we note that $\theta_{ij}\sim \mathcal O(L^{-2})$ within the region of support of the state $|\psi_n\rangle$, $S_n$. 
Hence $[e^{-i\theta_{ij}}-(1-i\theta_{ij})] \sim \mathcal O(L^{-4})$ for sites $i,j$ within $S_n$.
As a result, 
$
\norm{\delta H^{(2)}(t)|\psi_{n}\rangle} \sim \mathcal O(L^{-4}).
$
Thus, we obtain 
\begin{align}
\tilde \varepsilon _n- \varepsilon _n =& - \sum_b \theta_b   \int_{0}^T\!\!\frac{{\rm d}t}{T}\, \langle \psi_n|U^\dagger(t)\hat I_b(t) U(t)|\psi_n\rangle \notag \\
& +\mathcal O(L^{-5/2}).
\end{align}
Using that in a Floquet eigenstate the time-averaged expectation value  over one period equals the long-time average, we obtain
\begin{align}
\tilde \varepsilon _n- \varepsilon _n = - \sum_b \theta_b    \langle \psi_n|\bar I_b|\psi_n\rangle +\mathcal O(L^{-5/2}),
\end{align}
where $\bar{\mathcal O}$ denotes the long-time average in the Heisenberg picture [see Eq.~\eqref{eq:OperatorTimeAverageDef}].
Retracing the arguments in the main text that lead from Eq.~\eqref{eq:deltaepsilon} to Eq.~\eqref{eq:magdensres}, we find $\sum_b \theta_b \bar I_b  = \bar M B_0$. 
Thus, we conclude
\begin{align}
\tilde \varepsilon _n-  \varepsilon _n = -\langle \psi_n|\bar M |\psi_n\rangle B_0  +\mathcal O(L^{-5/2}).
\label{eqa:step_xyz}
\end{align}
This establishes Eq.~\eqref{eqa:eq00}, which was the first goal of this Appendix.

\subsection{Vanishing sum of corrections}
\label{seca:correction_sum}
As the final step of this Appendix, we now show that the corrections to Eq.~\eqref{eqa:step_xyz} (which individually scale with system size, $L$, as  $L^{-4}$), approximately cancel out when summed over all quasienergy levels in the $\ell$-particle subspace, yielding an  {\it exponentially} suppressed  correction: 
\begin{align}
\sum_{n}( \tilde \varepsilon _n - \varepsilon _n) =-  \sum_n &B_0\langle \psi_n|\bar M|\psi_n\rangle   +\mathcal O ( e^{-L/\xi_l}),
\label{eqa:eq11}
\end{align}

To establish Eq.~\eqref{eqa:eq11}, it is convenient to first express Eq.~\eqref{eqa:step_xyz} in terms of the magnetization densities on each plaquette, $\{\bar m_p\}$ by using $\bar M = \sum_p a^2 \bar m_p$:
\begin{align}
\tilde \varepsilon _n-  \varepsilon _n = - \sum_p a^2  B_0\langle \psi_n|\bar m_p |\psi_n\rangle   +\mathcal O(L^{-5/2}).
\label{eqa:step_xyz1234}
\end{align}
To obtain Eq.~\eqref{eqa:eq11} from the above result, we exploit the LIOM decomposition of the quasienergy levels in terms of the quasienergy coefficients $\varepsilon _{\alpha _1},\varepsilon _{\alpha _1\alpha _2},\ldots$ [Eq.~\eqref{eq:FloquetOperatorForm}], and  the analogous  decomposition time-averaged magnetization density in term of the magnetization coefficients $\bar m^p_{\alpha _1}, \bar m^p_{\alpha _1\alpha _2},\ldots$ [Eq.~\eqref{BarMpExpansion:eq:stab}].
By inserting these expansions into Eq.~\eqref{eqa:step_xyz1234} and using that Eq.~\eqref{eqa:step_xyz1234} holds for each Floquet eigenstate with up to $\ell$ particles (i.e., for each combination of up to $\ell$ excited  LIOMs), one can verify that, for each choice of $\ell$ LIOMs, $\al$,
\begin{align}
\tilde \varepsilon _\al- \varepsilon _\al = -B_0\sum_p a^2 \bar m^p_\al +\mathcal O(L^{-5/2}).
\label{eqa:step123_1}
\end{align}

We now seek to compute the sum the left hand side above  over all $\binom{D_1}{\ell}$ distinct combinations of $\ell$ LIOMs, where $\binom{a}{b}$ denotes the binomial coefficient and $D_1 = L^2/a^2$  the dimension of the system's single-particle subspace. 
Specifically, we seek to compute 
\be
\kappa_\ell \equiv \sum_{\al} (\tilde \varepsilon _\al- \varepsilon _\al ).
\ee
Since $\bar m^p_{\al}$ and $\varepsilon _\al$ may only be nonzero when the LIOMs $\al$ are located within a distance $\sim \xi_l$ from each other, there are of order $L^2/a^2$ combinations of $\ell$ LIOMs for which $\bar m^p_{\al}$ and $\varepsilon _\al$ may be significant.
Summing Eq.~\eqref{eqa:step123_1} over these $\mathcal O(L^2)$ combinations, we obtain
\begin{align}
\kappa_\ell =- B_0\sum_p \sum_\al a^2 \bar m^p_\al +\mathcal O(L^{-1/2}).
\label{eqa:step123_2}
\end{align}
To obtain $\kappa_\ell$, we use  $\kappa_1,\ldots \kappa_\ell$  to express the sum of $\tilde \varepsilon _n - \varepsilon _n$ over all $\ell$-particle quasienergy levels. 
An argument similar to the one made in Sec.~\ref{sec:cumulants} shows that the sum of $\tilde \varepsilon _n - \varepsilon _n$ over all $\ell$-particle quasienergy levels yields
\be 
\sum_n(\tilde \varepsilon _n - \varepsilon _n )  =\sum_{\ell'=1}^\ell \binom{D_1-\ell'}{\ell-\ell'} \sum_{\alpha_1 \ldots \alpha_{\ell'}} \kappa_\ell
\label{eqa:qe_kappa_relation}
\ee
Note, in particular, that  the sum of $\tilde \varepsilon _n-\varepsilon _n$ over all single-particle quasienergy levels is identical to $\kappa_1$. 
Using that $\sum_n(\tilde \varepsilon _n - \varepsilon _n ) $ must be quantized an integer multiple of $2\pi /T$ (see Sec.~\ref{ValuesOfTheInvariants:sec} in the main text) along with an inductive argument similar to the one below Eq.~\eqref{TrkMucorrespondence:eq:stab} in the main text,  we conclude that $\kappa_\ell$  must be an integer multiple of $2\pi/T$ for each $\ell \leq k$. 

Using inductive arguments similar to the ones employed above, using that $\Tr_{\ell'} \bar m_p = \Tr_{\ell'}\bar m_q$ for any two plaquettes $p,q$ in the lattice,  for any $\ell'=1,\ldots \ell$, it follows that, 
\begin{align}
  \sum_\al\bar m^p_\a =  \sum_\al \bar m^{q}_\al.
\label{eqa:step1234567}
\end{align}
Using this result in Eq.~\eqref{eqa:step123_2} along with $L^2 B_0 = 2\pi$, we thus find, for any given plaquette $p_0$ in the lattice,
\begin{align}
\kappa_\ell = 2\pi \sum_\al \bar m^{p_0}_\al +\mathcal O(L^{-1/2}).
\label{eqa:step123456}
\end{align}

We now consider how the right- and left-hand sides differ from their values in the thermodynamic limit, $L\to \infty$.
Firstly, $ \bar m^{p_0}_\al$ is exponentially suppressed for all but the $~\binom{\xi_l^2/a^2}{\ell}$ choices of  $\ell$ LIOMs  where all LIOM  centers are all located  within a distance $\sim \xi_l$ from plaquette $p_0$. 
Hence $ \sum_\al \bar m^{p_0}_\al$ only depends on the details of the system near plaquette $p$, and therefore can only differ from its value  in the thermodynamic limit by an amount of order $e^{-L/\xi_l}$. 
From below Eq.~\eqref{eqa:qe_kappa_relation} we recall that $\kappa_\ell$ is exactly quantized as an integer multiple of $2\pi /T$.
Moreover, Eq.~\eqref{eqa:step123456}  shows that $\kappa_\ell$ can only differ from  $ \sum_\al \bar m^{p_0}_\al$ by an amount of order $\mathcal O(L^{-2})$. 
Hence, when $L\gg \xi_l$, $\kappa_\ell$ must be exactly identical to its value in thermodynamic limit. 
We conclude that   $\delta_\ell \equiv \kappa_\ell - 2\pi \sum_\al \bar m^{p_0}_\al $ can only differ from its value in the thermodynamic limit by an amount of order $e^{-L/\xi_l}$. 
Since $\delta_\ell = 0$ in the thermodynamic limit, we find, for each plaquette in the system, $p_0$,
\begin{align}
\sum_{\al} (\varepsilon _\al -  2\pi \bar m^{p_0}_\al)  =  \mathcal O(e^{-L/\xi_l}).
\label{eqa:step7}
\end{align}
Using  $\bar M =\sum_p a^2 \bar m_p$ along with the LIOM decompositions in Eqs.~\eqref{eq:FloquetOperatorForm}~and~\eqref{BarMpExpansion:eq:stab}, we conclude that 
Eq.~\eqref{eqa:eq11} holds.
This was the goal of this subsection, and concludes this Appendix.

\end{document}